\documentclass[a4paper]{jpconf}
\RequirePackage{lineno} 
\usepackage{graphicx}
\usepackage{float} 
\usepackage{amssymb}
\usepackage{cite} 
\usepackage{harpoon}
\usepackage{hyperref}
\usepackage{siunitx}
\hypersetup{linkcolor=blue,citecolor=blue,filecolor=black,urlcolor=blue}
\newcommand{\pp}{\mbox{$p+p$}}
\newcommand{\pPb}{\mbox{$p$+Pb}}
\newcommand{\pta}{\mbox{$p_{\mathrm{T}}^{\mathrm{a}}$}}
\newcommand{\ptb}{\mbox{$p_{\mathrm{T}}^{\mathrm{b}}$}}
\newcommand{\ptab}{\mbox{$p_{\mathrm{T}}^{\mathrm{a,b}}$}}

\newcommand{\PbPb}{\mbox{Pb+Pb}}
\newcommand{\invmub}{\mbox{$\mu{\rm b}^{-1}$}}
\newcommand{\nchave}{\mbox{$\langle N_{\mathrm{ch}}\rangle$}}
\newcommand{\signch}{\mbox{$\sigma_{\tiny{N_{\mathrm{ch}}}}$}}

\newcommand{\nch}{\mbox{$N_{\mathrm{ch}}$}}
\newcommand{\nchrec}{\mbox{$N_{\mathrm{ch}}^{\mathrm{rec}}$}}

\newcommand{\ETfcal}{\mbox{$\Sigma E_{\mathrm{T}}^{{\scriptscriptstyle \mathrm{Pb}}}$}}
\newcommand{\ETfcalave}{\mbox{$\langle\Sigma E_{\mathrm{T}}^{{\scriptscriptstyle \mathrm{Pb}}}\rangle$}}

\newcommand{\sqrtsnn}{\mbox{$\sqrt{s_{\mathrm{NN}}}$}}
\newcommand{\Dphi}{\mbox{$\Delta \phi$}}
\newcommand{\Deta}{\mbox{$\Delta \eta$}}

\newcommand{\Yphi}{\mbox{$Y(\Delta\phi)$}}

\newcommand{\DelYint}{\mbox{$\Delta Y_{\mathrm{int}}$}}
\newcommand{\pT} {\ensuremath{p_{\mathrm{T}}}}
\begin{document}
\title{Long-range correlations in proton-lead collisions at $\sqrtsnn = 5.02$~TeV from ATLAS}

\author{Jiangyong Jia on behalf of the ATLAS Collaboration }
\address{Chemistry Department, Stony Brook University, Stony Brook, NY 11794, USA}
\address{Physics Department, Brookhaven National Laboratory, Upton, NY 11796, USA}

\ead{jjia@bnl.gov}
\begin{abstract}
Two-particle correlations in relative azimuth $\Delta\phi$ and relative pseudorapidity $\Delta\eta$ are studied in $\pPb$ collisions at $\sqrt{s_{\mathrm{NN}}}=5.02$ TeV with the ATLAS detector at LHC. The correlations are studied as a function of charged particle $p_{\mathrm T}$ and the collision $\ETfcal$ summed over $3.1 < \eta < 4.9$ in the direction of the Pb beam. After subtracting the known sources of correlations such as dijets, resonances and momentum conservation, estimated using events with low $\ETfcal$, the resulting correlations exhibit a $\Delta\phi$ modulation that is flat in $\Delta\eta$ out to $|\Delta\eta|=5$. The modulation is comparable in magnitude to similar modulations observed in heavy ion collisions, and can be described by a $1+2c_2\cos2\Delta\phi+2c_3\cos3\Delta\phi$ function over $0.5<\pT<7$ GeV in broad ranges of $\ETfcal$. The correlation analysis is repeated for event classes defined by the number of reconstructed charged particles $N_{\mathrm{ch}}^{\mathrm{rec}}$. This analysis gives nearly the same result as the analysis based on $\ETfcal$ for the long-range correlation at the near-side ($\Delta\phi\sim0$), but leads to biases in the long-range correlations at the away-side ($\Delta\phi\sim\pi$). HIJING simulation suggests that this bias is mainly associated with the contributions from dijets which are correlated strongly with the $N_{\mathrm{ch}}^{\mathrm{rec}}$.
\end{abstract}
\section{Introduction}
\begin{figure}[!t]
\centering
\includegraphics[width=1\columnwidth]{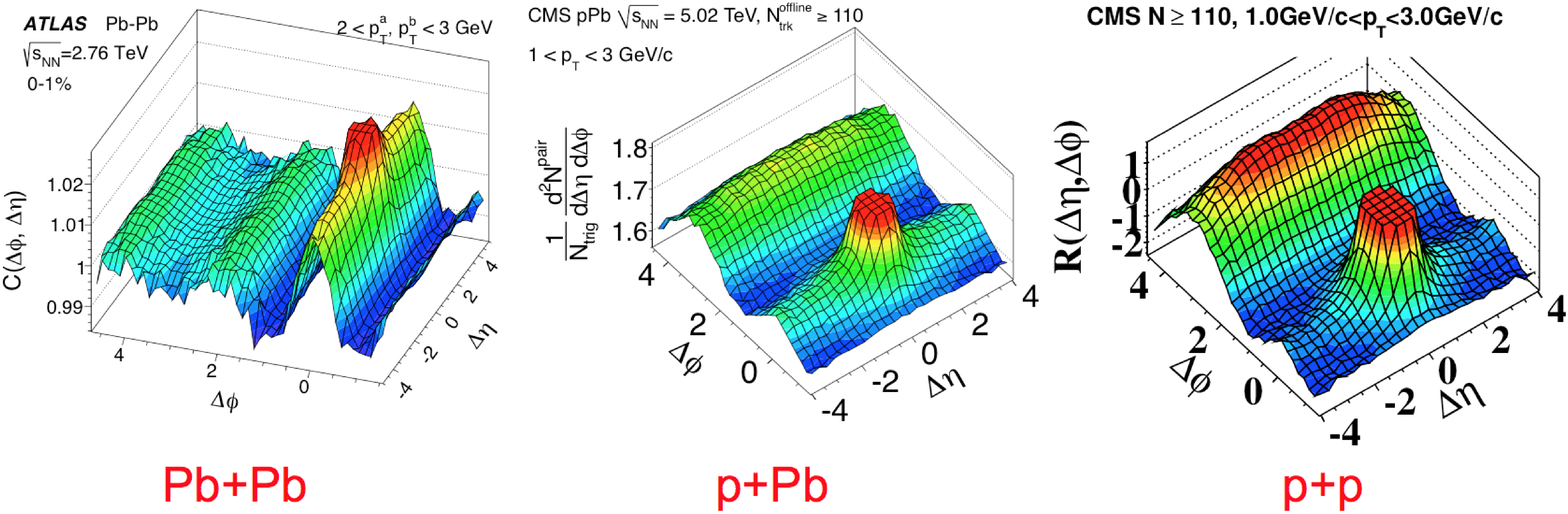}
\caption{\label{fig:intro} Examples of two-particle correlations in $\Dphi$ and $\Deta$ in $\PbPb$~\cite{ATLAS:2012at}, $\pPb$~\cite{CMS:2012qk} and $\pp$~\cite{Khachatryan:2010gv} collisions.}
\end{figure}

Recent studies of two-particle angular correlations (2PC) in relative azimuthal angle, $\Delta\phi=\phi_a-\phi_b$, and relative pseudorapidity, $\Delta\eta=\eta_a-\eta_b$, in $\pp$, $p$+A and A+A collisions have generated considerable interest in both high energy and heavy-ion physics communities. In events with very large multiplicity, an enhanced production of particle pairs is observed at small relative azimuthal angle $\Delta\phi\sim0$, and this excess extends over a long-range in $\Delta\eta$, i.e. extends to at least $|\Delta\eta|=5$. This so called ``ridge'' phenomena was first observed in Au+Au or $\PbPb$ collisions at RHIC and LHC~\cite{Abelev:2009af,Alver:2009id,Adare:2008cqb,ATLAS:2012at,Aamodt:2011by,Chatrchyan:2012wg}, but was found to also exist in high-multiplicity $\pp$ collisions~\cite{Khachatryan:2010gv} and $\pPb$ collisions~\cite{CMS:2012qk,ALICE:2012,Aad:2012gla} (see Fig.~\ref{fig:intro}). The ``ridge'' is a reflection of QCD dynamics in high-density system, where many hadrons are produced in a small volume. However, the effective mechanism may depend on the system size. In heavy-ion collisions, the ridge is commonly believed to be the result of collective flow of the produced matter in the final state. That flow has been successfully described by the relativistic hydrodynamic calculations~\cite{Voloshin:2008dg}. In high-multiplicity $\pp$ and $\pPb$ collisions, however, the mechanism responsible for the ridge is not yet clear. Some models interpret the ridge as resulting from finial-state effects via the hydrodynamic picture~\cite{Bozek:2011if,Bozek:2012gr,Shuryak:2013ke,Bozek:2013uha}, while others argue that the ridge is due to initial-state gluon saturation effects~\cite{Dusling:2012wy,Dusling:2013oia}.

One striking feature of the long-range correlation in Au+Au/$\PbPb$ collisions is that the ridge is not only restricted to $\Delta\phi\sim0$ (near-side), but also appears around $\Delta\phi\sim\pi$ (away-side). The detailed shape of the ridge in $\Delta\phi$ depends on the $\pT$ selection of particles in the pair, and has been described in the framework of harmonic flow~\cite{ATLAS:2012at,Aamodt:2011by,Chatrchyan:2012wg}. If the ridge in high-multiplicity $\pp$ or $\pPb$ collisions is due to a similar mechanism as that responsible for the ridge in large systems such as $\PbPb$, one would also expect the presence of a ridge at the away-side. The extraction of the away-side ridge in $\pp$ or $\pPb$ collisions, however, is complicated by the correlations arising from known sources, such as dijets, resonance and momentum conservation, collectively referred to as ``recoil''~\cite{ATLAS:2012at}. In a recent publication~\cite{Aad:2012gla}, ATLAS has developed a data-driven procedure to separate the contribution from away-side recoil and the away-side long-range correlations in $\pPb$ collisions. This subtraction permits a detailed study of the long-range correlation over the full $\Delta\phi$ range, as well as the extraction of the associated harmonic coefficients as a function of $\pT$, $\eta$, and event multiplicity or activity. 

This proceedings provides details of the ATLAS $\pPb$ ridge results at Ref.~\cite{Aad:2012gla} with an extended set of plots given at Ref.~\cite{atlaslink}. We first describe how ATLAS classifies the activity of the events, and then discuss the recoil-removal procedure and explain why it can be used to expose the away-side ridge in $\pPb$ collisions. We review the properties of the ridge as a function of event multiplicity, $\pT$, $\Delta\eta$ and charge combination. We present the multiplicity and $\pT$ dependence of the harmonic coefficients obtained via a Fourier analysis of the $\Dphi$ distribution. Finally, this proceedings comments on potential auto-correlation biases associated with different definitions of event classes.
\section{Event and track selections}
\label{sec:1}
\begin{figure}[!t]
\centering
\includegraphics[width=0.5\columnwidth]{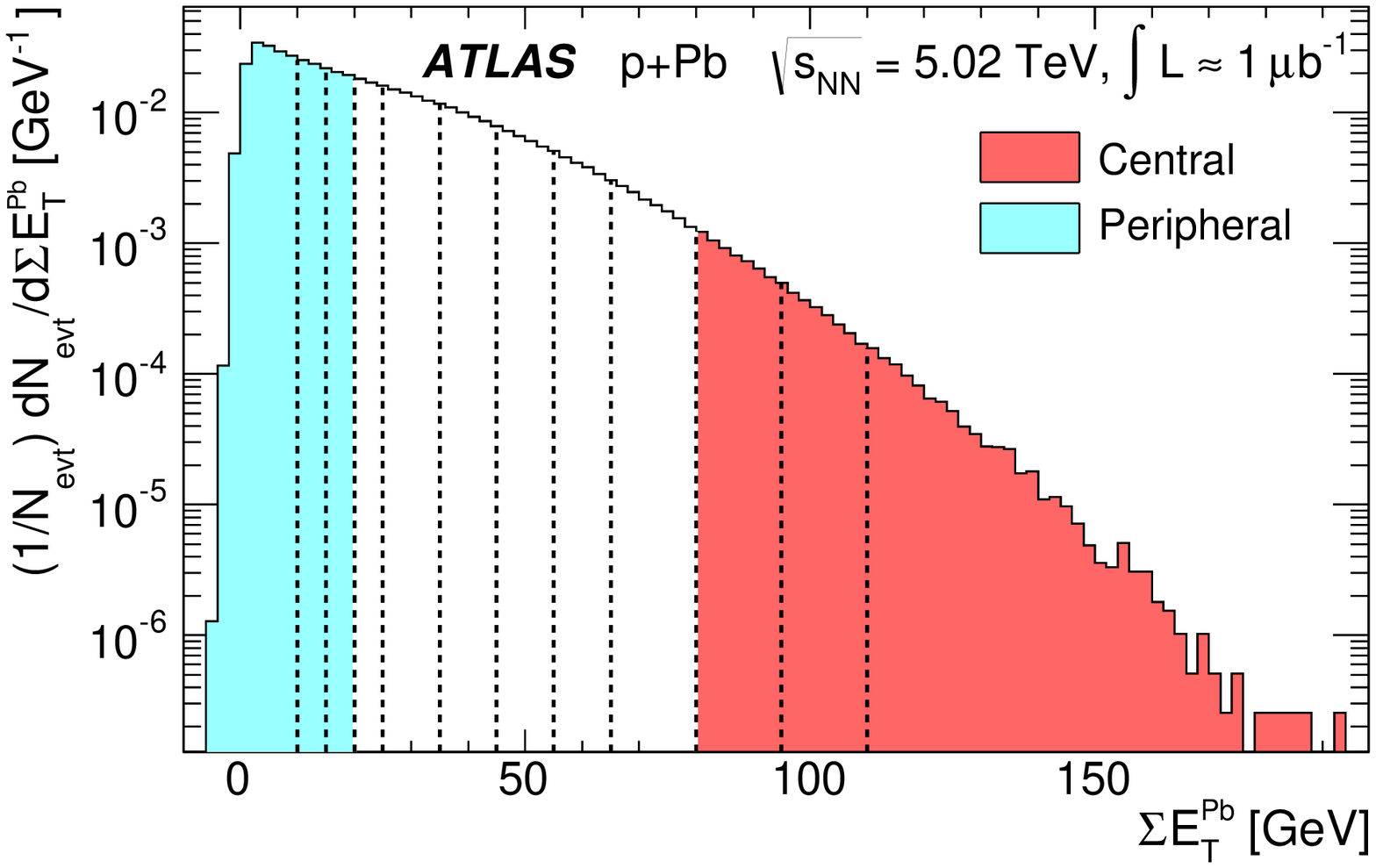}\includegraphics[width=0.5\columnwidth]{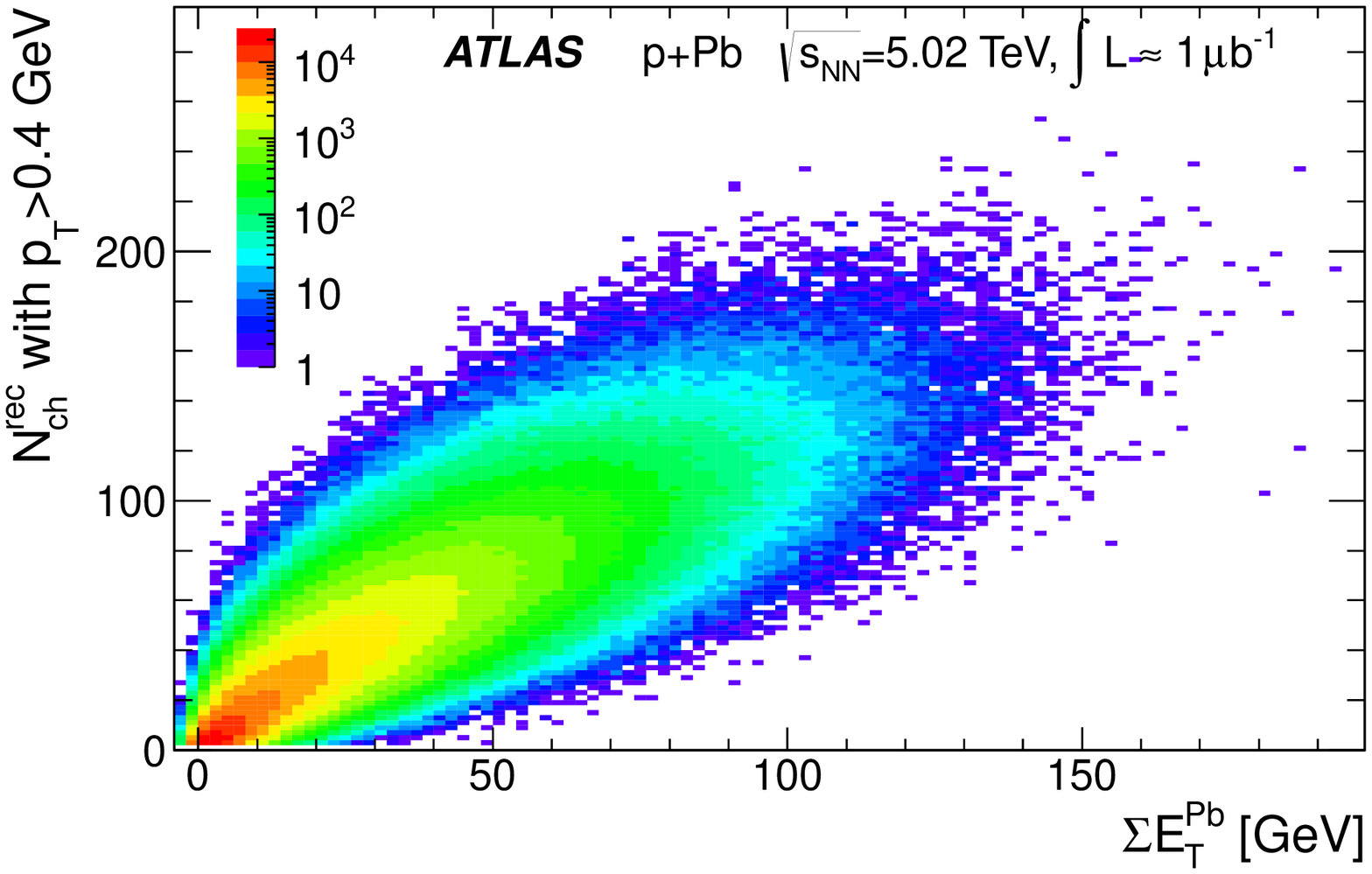}
\caption{\label{fig:cent} (left) Distribution of \ETfcal\ for minimum-bias $\pPb$ events. Vertical lines indicate the boundaries of the event activity classes~\cite{Aad:2012gla}. (right) Correlation between $\ETfcal$ and number of reconstructed charged particles with $\pT>0.4$ GeV and $|\eta|<2.5$ in the same events~\cite{atlaslink}.}
\end{figure}

The result is based on $\pPb$ collisions corresponding to 1~$\invmub$, or about 2 million events, at $\sqrtsnn = 5.02$~TeV recorded during a short run in September 2012. The 2PC analysis is based on charged particle tracks with $\pT>0.4$~GeV and $|\eta|<2.5$, reconstructed by the ATLAS inner detector~\cite{Aad:2008zzm} with an overall efficiency of about 70\%. The events are divided into 11 narrow activity classes based on the total transverse energy, $\ETfcal$, measured by the forward calorimeter over $3.1<\eta<4.9$ in the Pb-going direction (see Fig.~\ref{fig:cent}). Four larger intervals, \mbox{$\ETfcal > 80$~GeV},  \mbox{$55<\ETfcal < 80$~GeV}, \mbox{$25<\ETfcal < 55$~GeV} and \mbox{$\ETfcal < 20$~GeV}, are used for detailed studies of the 2PC as a function of $\pT$. Event activity classes are also defined by selecting on the charged particle track multiplicity $\nch$ in the detector. However, this approach is found to introduce auto-correlation bias to the 2PC measurements (see Sec.~\ref{sec:5}).  On the other hand, the average value $\langle\nch\rangle$, calculated for each $\ETfcal$ event class, would have little auto-correlations with the 2PC measurements, and they are given in Table~\ref{tab:1}.

\begin{table*}[h]
\centering\footnotesize
\begin{tabular}{|c |c|c|c|c|c|c|c|c|}\hline

\ETfcal [GeV] &$>110$ &95-110 &80-95 &65-80 & 55-65 &45-55&35-45 &25-35\tabularnewline
Fraction [\%]     &0.21  &0.45 & 1.24 & 3.11 &3.99 & 6.37&9.71  & 13.80 \tabularnewline
\ETfcalave [GeV]    &122.4 &101.2&86.4  & 71.4 &59.6 & 49.7&39.7  & 29.7  \tabularnewline
\nchave             &183$\pm$8&160$\pm$7&141$\pm$6&123$\pm$6&107$\pm$5&93.3$\pm$4.2&78.8$\pm$3.6&63.3$\pm$2.9\tabularnewline
$\signch$           &37.0$\pm$2.1 &33.1$\pm$1.9 &31.5$\pm$1.8 &29.6$\pm$1.7 &27.6$\pm$1.6 &25.9$\pm$1.5 &24.1$\pm$1.4&21.8$\pm$1.2\tabularnewline\hline
\ETfcal [GeV] &20-25 &15-20 &10-15  &$<10$&$>80$ &55-80  &25-55 &$<20$ \tabularnewline
Fraction [\%]     &8.67 & 10.11 &11.98 &30.36&1.90  & 13.47 &29.88 & 52.45\tabularnewline
\ETfcalave [GeV]    &22.4 & 17.4 &12.4   &4.9&94.4  & 64.8  &37.3 & 9.0\tabularnewline
\nchave             &51.0$\pm$2.3&41.8$\pm$1.9&31.7$\pm$1.5&15.9$\pm$0.7&150$\pm$7&114$\pm$5&74.7$\pm$3.4&24.5$\pm$1.1\tabularnewline
$\signch$           &19.6$\pm$1.1&17.9$\pm$1.0&15.7$\pm$0.9&11.8$\pm$0.7&35.2$\pm$2.0 &29.4$\pm$1.7 &26.1$\pm$1.5&17.5$\pm$1.0\tabularnewline\hline
\end{tabular}\normalsize\vspace*{-0.2cm}
\caption{\label{tab:1} A list of $\ETfcal$ classes, the associated percentage of events and average value \ETfcalave. For each class, the efficiency corrected average number of charged particles with $\pT>0.4$ GeV and $|\eta|<2.5$ \nchave, standard deviation $\signch$, and their systematic uncertainties are also listed~\cite{atlaslink}. Here $\signch$ is the estimate of the physical fluctuation of $\nch$ for events selected in a given \ETfcal\ range, the spread of $\nch$ due to detector inefficiency is much smaller and has been subtracted.}\vspace*{-0.4cm}
\end{table*}

\section{Correlation function and recoil subtraction}
\label{sec:2}
The two-particle correlation function is constructed as the ratio of the distribution for pairs in the same-event and mixed-event:
\begin{eqnarray} 
\label{eq:1}
C(\Dphi,\Deta) = \frac{S(\Dphi,\Deta)}{B(\Dphi,\Deta)}\;,\;\; C(\Dphi) =\frac{S(\Dphi)}{B(\Dphi)},
\end{eqnarray}
with each particle weighted by the inverse of the reconstruction efficiency. The mixed-pair distribution $B(\Dphi, \Deta)$ is designed to remove residual structures due to detector acceptance and occupancy from the $S(\Dphi,\Deta)$. The 1D distributions $S(\Dphi)$ and $B(\Dphi)$ are obtained by integrating $S(\Dphi,\Deta)$ and $B(\Dphi,\Deta)$, respectively, over $2<|\Delta\eta|<5$. The normalization of $C(\Dphi,\Deta)$ is chosen such that the \Dphi-averaged value of $C(\Dphi)$ is unity.

Denoting $N_a$ and $N_b$ as the average number of trigger and associated particles per-event for given $\pT$ selection, the meaning of the correlation function can be understood as follows. For a perfect detector, the mixed-event distribution describes the rate of combinatorial pairs $\pi B = N_aN_b$, while the same event pairs contains both the correlated pairs and combinatorial pairs, i.e. $\pi S = J(\Dphi) + \xi_{_\mathrm{ZYAM}} N_aN_b$. Factor $\xi_{_\mathrm{ZYAM}}$ reflects the magnitude of a flat pedestal, estimated via the Zero-Yield at Minimum (ZYAM) procedure~\cite{Ajitanand:2005jj,Adare:2008cqb}. Hence $C(\Dphi) = J(\Dphi)/(N_aN_b)+\xi_{_\mathrm{ZYAM}}$, and the correlated component is measured relative to the underlying event background.

\begin{figure}[!t]
\centering
\includegraphics[width=0.7\columnwidth]{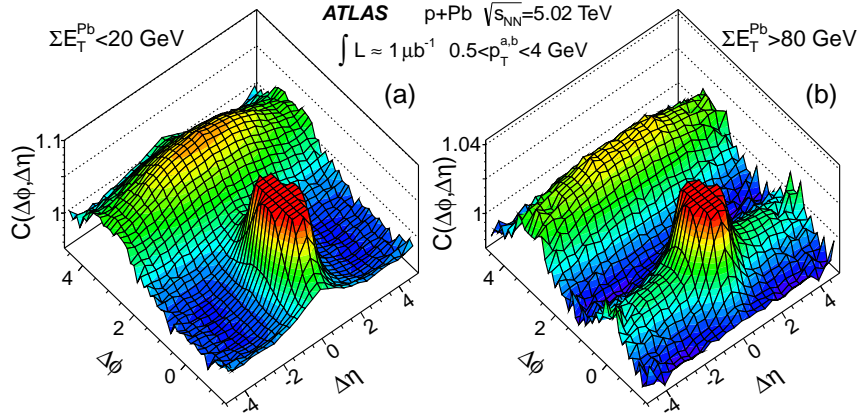}
\caption{\label{fig:m1a} Two-dimensional correlation functions for peripheral events and central events, both with a truncated maximum to suppress jet fragmentation peak around $(\Delta\eta,\Delta\phi)=(0,0)$~\cite{Aad:2012gla}.}
\end{figure}

Examples of the 2D correlation function are shown in Fig.~\ref{fig:m1a}. A clear long-range component in the right panel, on the order of a few percent of the background, is seen in the near-side. A similar long-range correlation, less flat in $\Deta$, is also seen in the away-side. However, a significant fraction of this away-side correlation is due to recoil to the trigger particle, which complicates the extraction of the genuine long-range correlation.

The effects of recoil or momentum conservation, in a system with finite number of particles, have been studied previously~\cite{Borghini:2000cm,Borghini:2002mv}. Under very general assumptions, these effects were shown to give a $\cos (\Dphi)$ correction to the correlation function:
\begin{eqnarray}
\label{eq:recoil}
\delta C(\Dphi) = -\frac{\pT^{\mathrm a}\pT^{\mathrm b}}{N\langle \pT^2\rangle} \cos(\Dphi) \propto -\frac{\pT^{\mathrm a}\pT^{\mathrm b}}{N_b\langle \pT^2\rangle}  \cos(\Dphi)
\end{eqnarray}
where $N$ is the average particle multiplicity, which is proportional to the average number of charged particles $N_b$ in a given associated $\pT$ range (typically 0.5-4 GeV in this analysis). In order to remove the recoil effects in $\pPb$ collisions, ATLAS uses the per-trigger yield defined as:
\begin{eqnarray}
\label{eq:yld}
Y(\Dphi)=\left( \frac{\int B(\Dphi) d\Dphi}{\pi N_{a}} \right) C(\Delta \phi) - b_{_{\mathrm{ZYAM}}} 
\end{eqnarray}
For an ideal detector, $\int B(\Dphi) d\Dphi=N_aN_b$, and Eq.~\ref{eq:yld} reduces to
\begin{eqnarray}
\label{eq:recoil1}
Y(\Dphi) = N_{b}\left(C(\Delta \phi) -\xi_{_{\mathrm{ZYAM}}}\right) =N_{b}C(\Delta \phi) -b_{_{\mathrm{ZYAM}}}, b_{_{\mathrm{ZYAM}}}= N_{b}\xi_{_{\mathrm{ZYAM}}}\;.
\end{eqnarray}
Hence the contribution of the recoil in Eq.~\ref{eq:recoil} to the per-trigger yield is:
\begin{eqnarray}
\label{eq:recoil2}
\delta Y(\Dphi) = N_{b} \delta C(\Delta \phi) \propto - \frac{\pT^{\mathrm a}\pT^{\mathrm b}}{\langle \pT^2\rangle}\cos(\Dphi).
\end{eqnarray}
So as long as the values of $\langle \pT^2\rangle$ are similar between different event classes, the recoil contribution can be estimated from peripheral events and subtracted from the central events. 

\begin{figure}[h]
\includegraphics[width=0.33\linewidth]{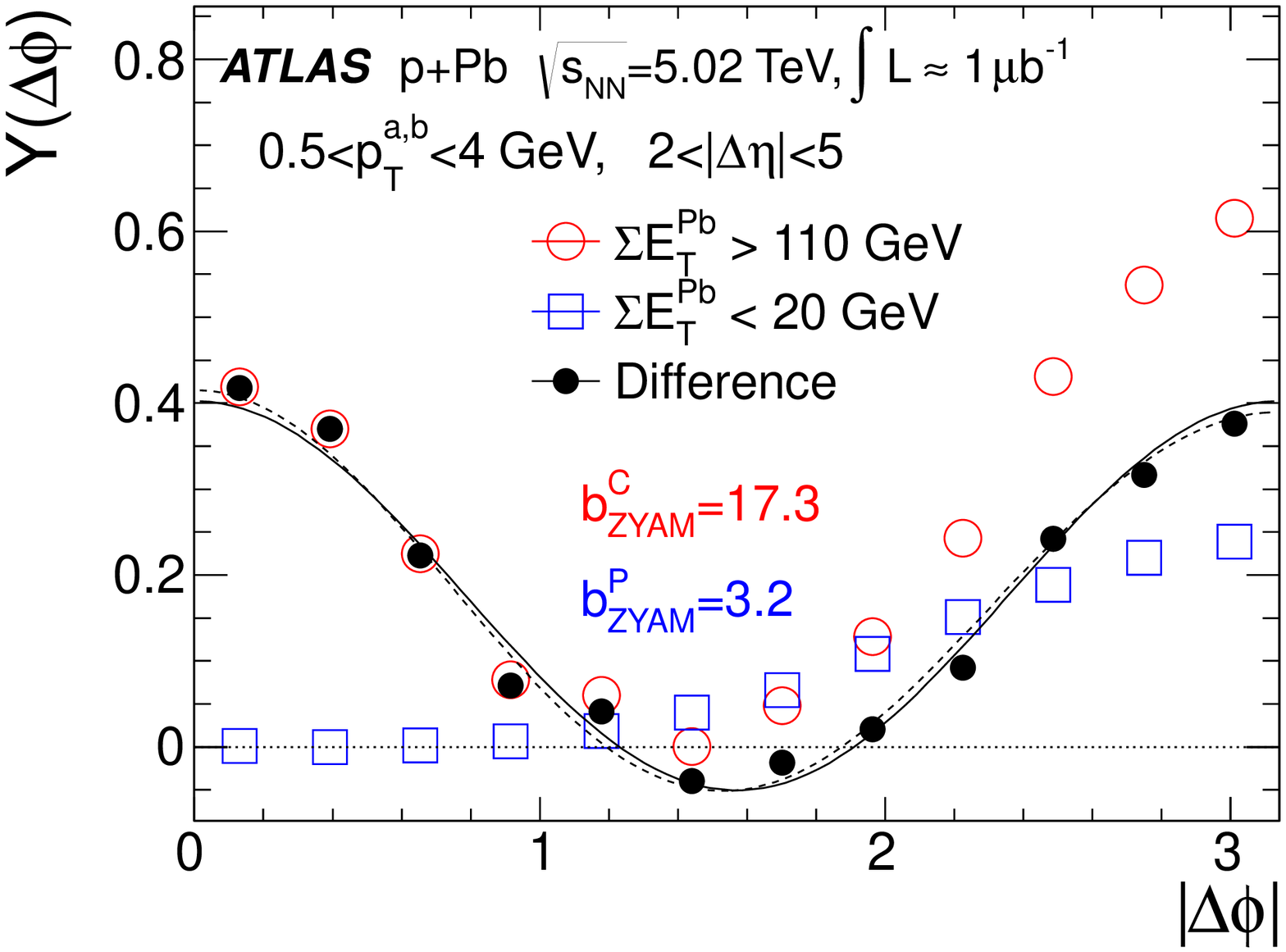}\includegraphics[width=0.33\linewidth]{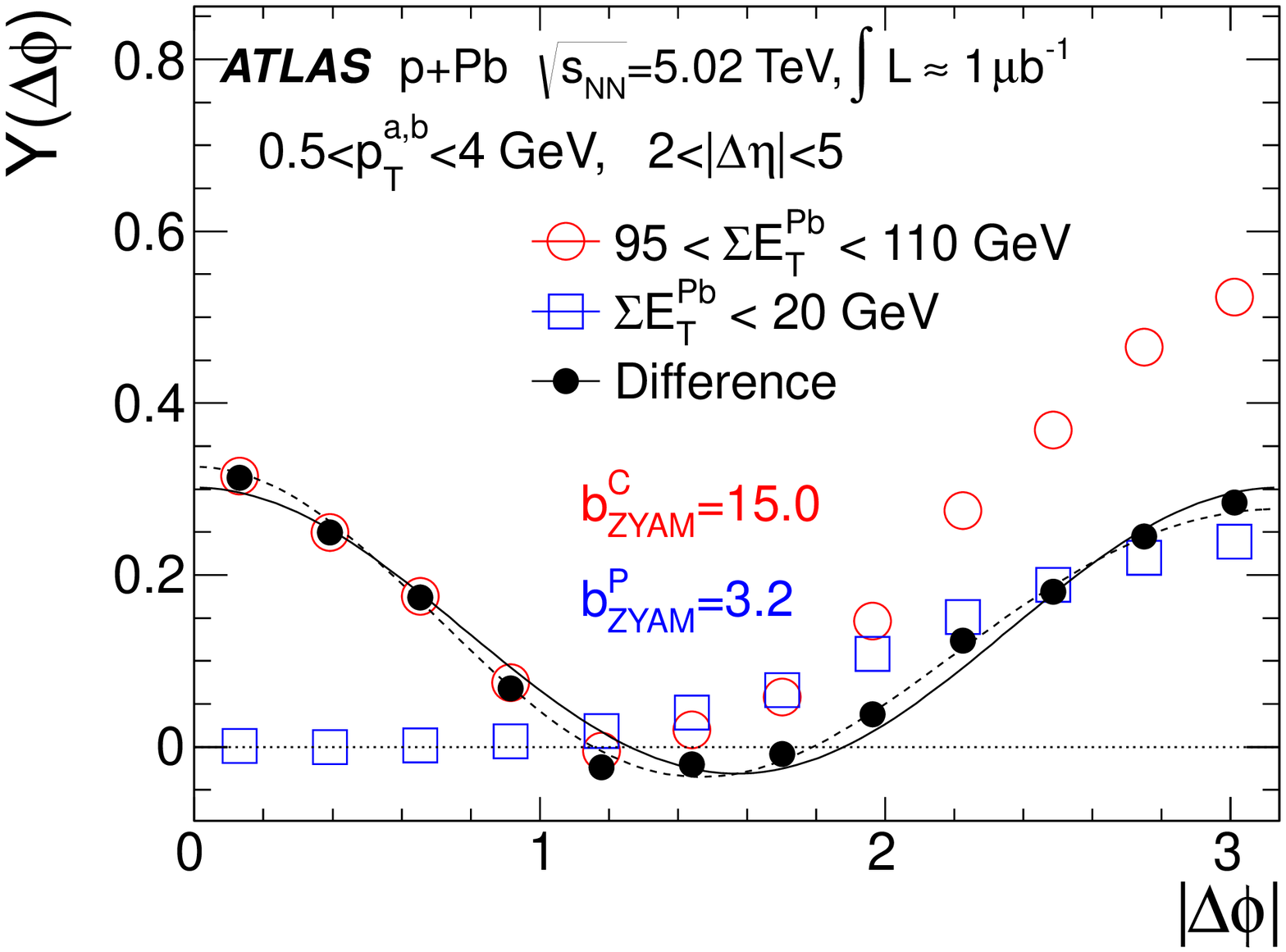}\includegraphics[width=0.33\linewidth]{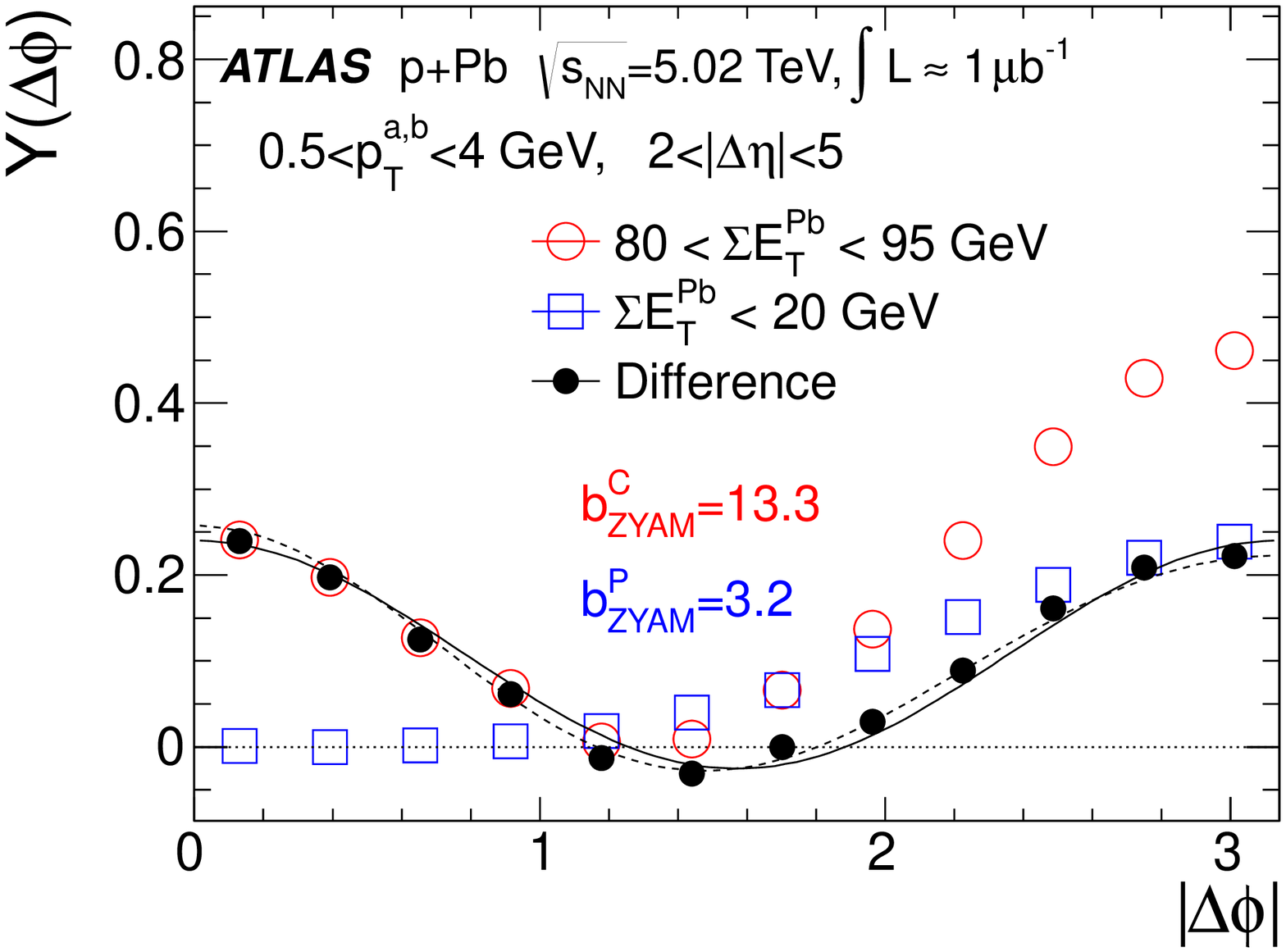}\\
\includegraphics[width=0.33\linewidth]{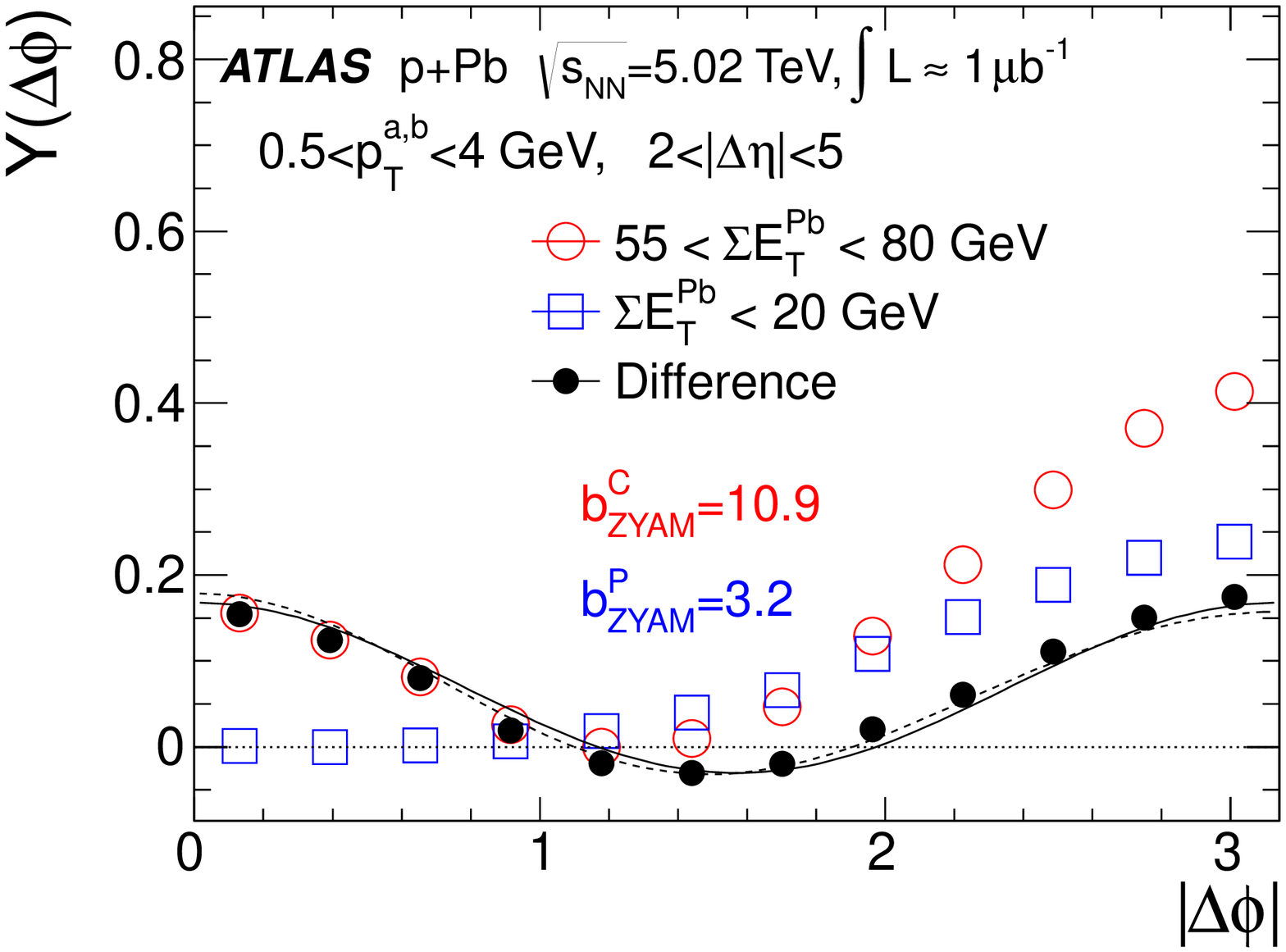}\includegraphics[width=0.33\linewidth]{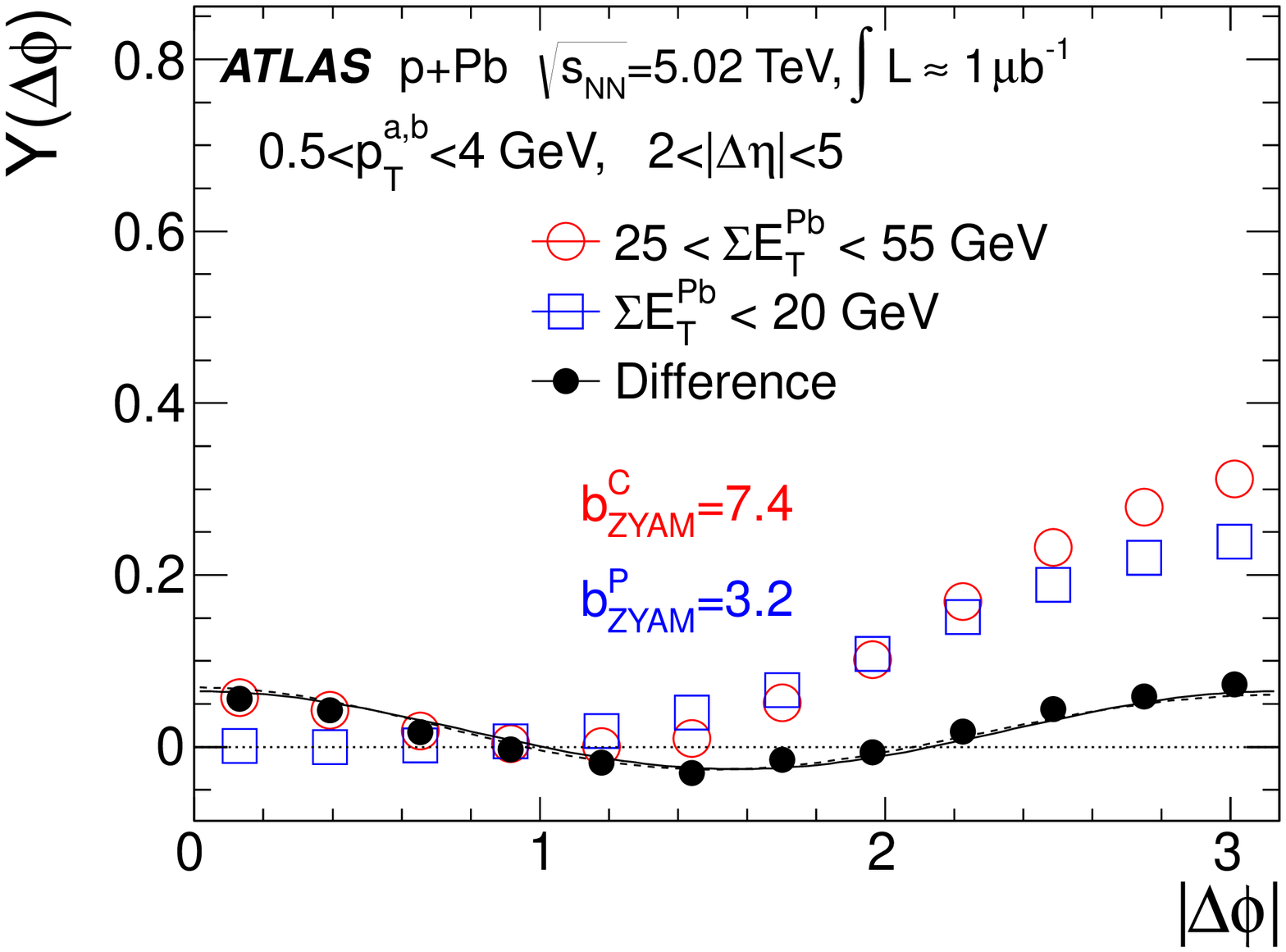}
\caption{\label{fig:m1b} Distributions of per-trigger yield in a given \ETfcal\ class (open circles), the $\ETfcal<20$ GeV class (open boxes), and their difference (solid circles), together with functions,  $a_0+2a_2\cos2\Delta\phi$ (solid lines) and $a_0+2a_2\cos2\Delta\phi+2a_3\cos3\Delta\phi$ (dashed lines), obtained via a Fourier decomposition~\cite{atlaslink}. The values for the ZYAM-determined pedestal levels are indicated on each panel for peripheral ($b^{\mathrm{P}}_{_{\mathrm{ZYAM}}}$) and central ($b^{\mathrm{C}}_{_{\mathrm{ZYAM}}}$)  \ETfcal\ bins.}
\end{figure}

Figure~\ref{fig:m1b} overlays the per-trigger yield in various $\ETfcal$ classes with that for $\ETfcal<20$ GeV. The latter has an approximately $-\cos\Dphi$ shape, suggesting that it reflects mainly the recoil effects. The subtracted distributions, $\Delta\Yphi$, obtained as:
\begin{eqnarray}
\label{eq:recoil2b}
\Delta Y(\Dphi) = Y(\Dphi)|_{\mathrm{a\;\; given}\;\; \ETfcal\;\;\mathrm{range}}-Y(\Dphi)|_{\ETfcal<20 \mathrm{GeV}},
\end{eqnarray}
are nearly symmetric around $\Dphi = \pi/2$ and can be well described by a $a_0 + 2a_2 \cos{2\Dphi}$ function. Including a $2a_3 \cos{3\Dphi}$ term only slightly improves the agreement with the data. This observation indicates that the long-range component of the two-particle correlations can be approximately described by a recoil contribution plus a $\Dphi$-symmetric ``double-ridge'' component. The magnitude of the double-ridge is comparable to the recoil contribution for events with $\ETfcal>55$ GeV (top 10\% of events according to Table~\ref{tab:1}), but increases to about twice the recoil contribution for events with $\ETfcal>110$ GeV (top 0.2\% of events).


\section{Properties of the ``double-ridge''}
\label{sec:3}
\begin{figure}[!h]
\centering
\includegraphics[width=0.75\columnwidth]{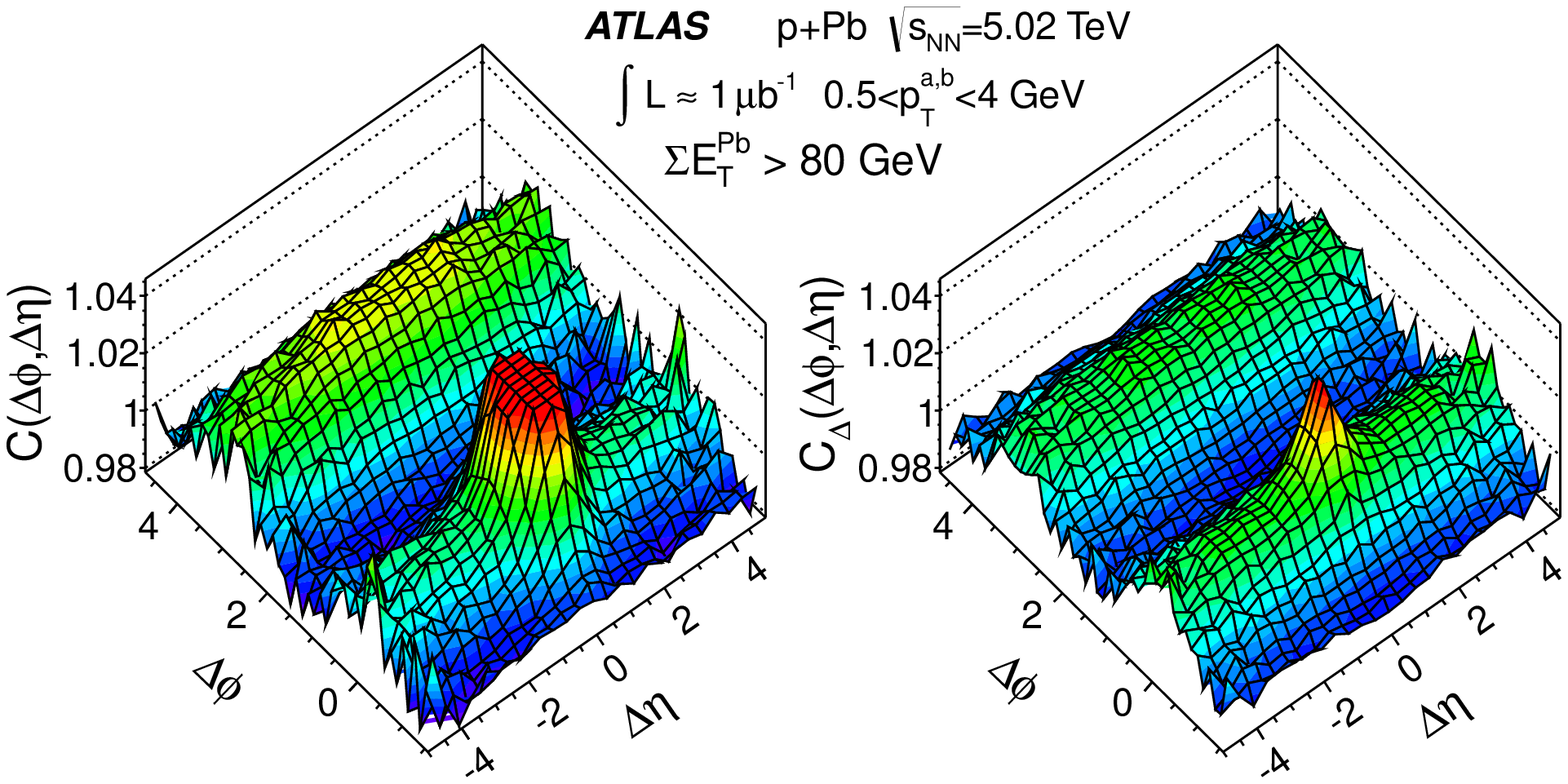}\vspace*{-0.2cm}
\includegraphics[width=0.75\columnwidth]{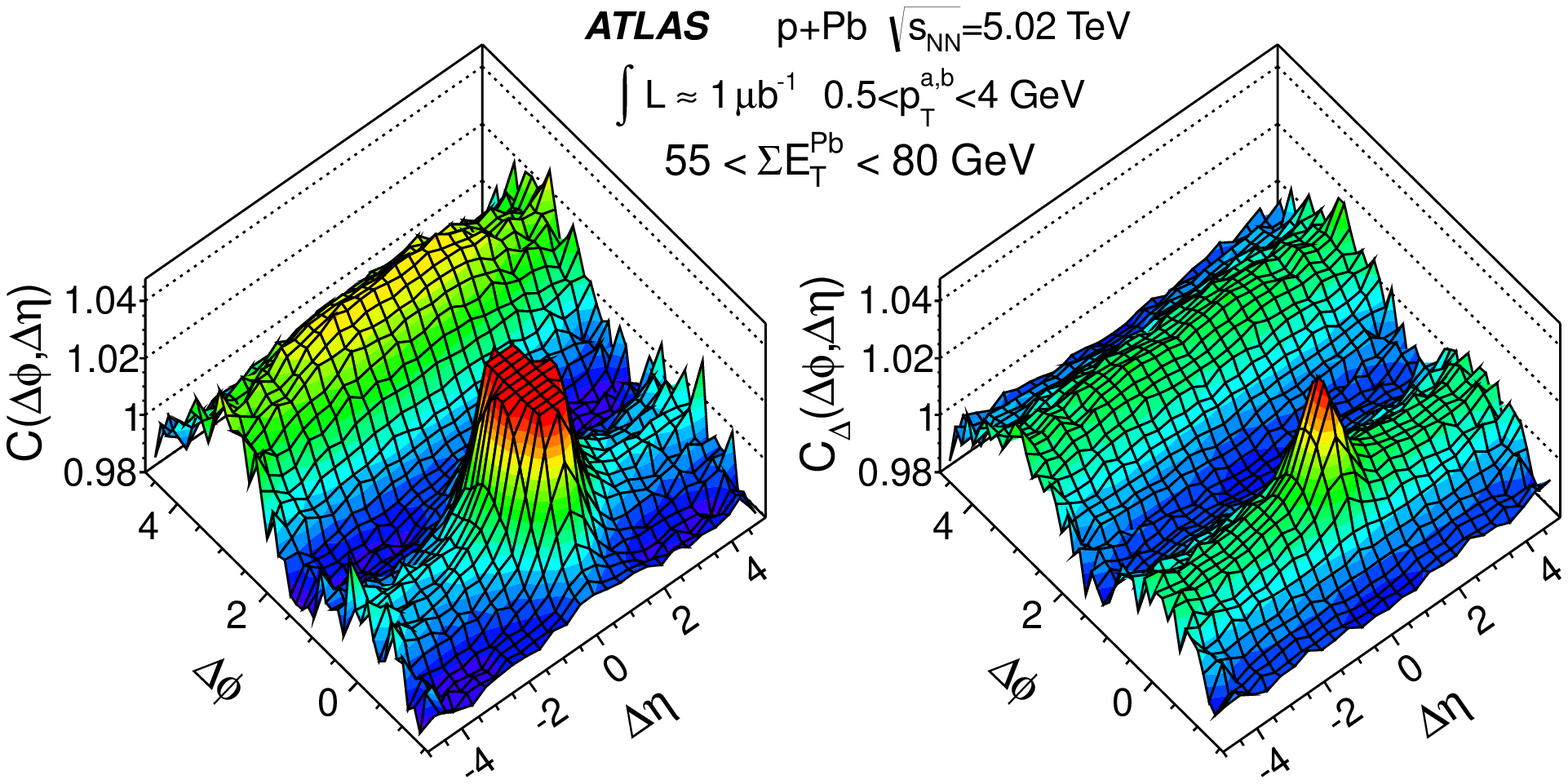}\vspace*{-0.2cm}
\includegraphics[width=0.75\columnwidth]{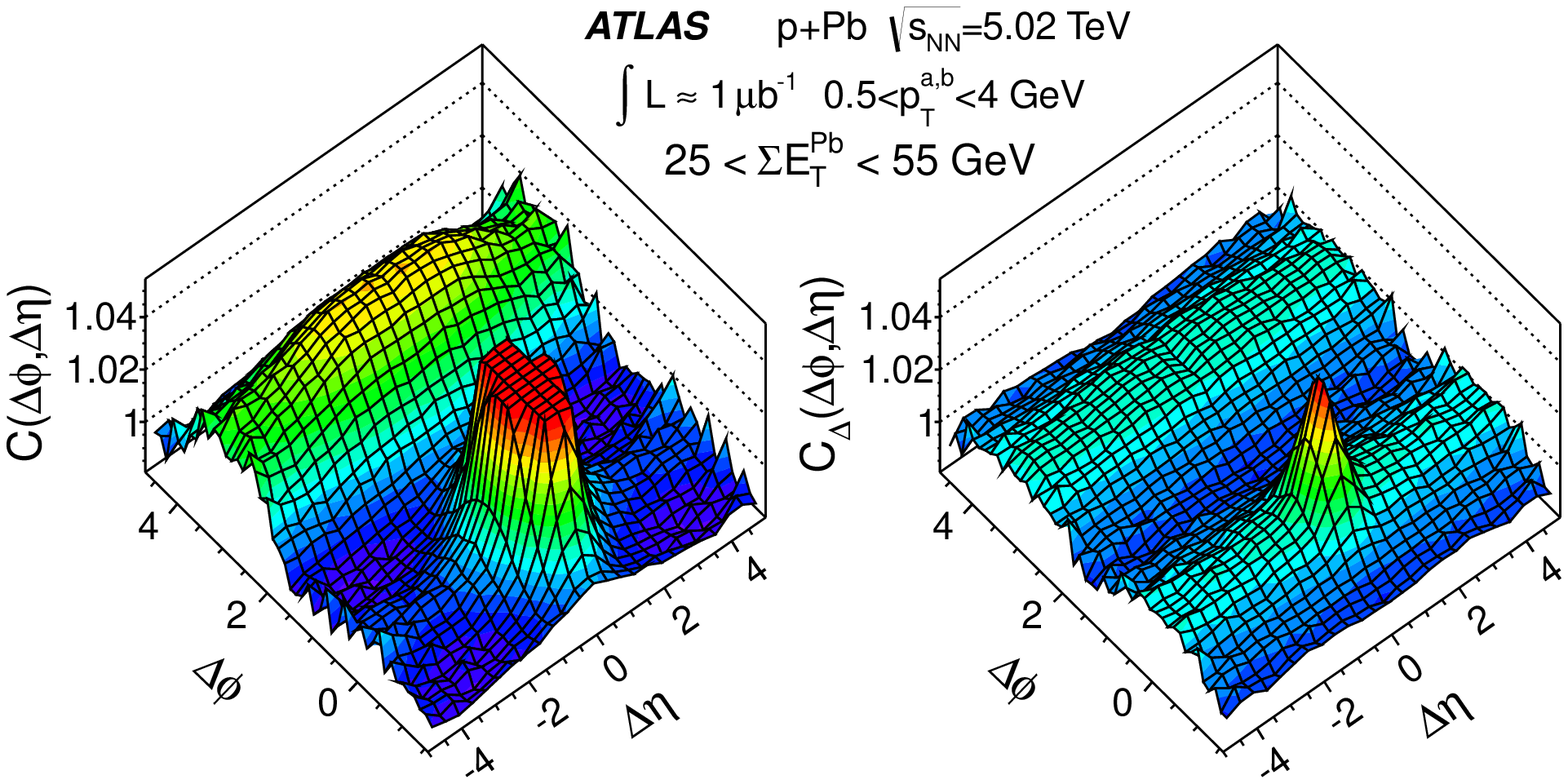}\vspace*{-0.2cm}
\caption{\label{fig:m2a} Two-dimensional correlation function, $C(\Delta\phi,\Delta\eta)$, without (left) and with (right) subtraction of recoil contribution in events with $\ETfcal>80$ GeV (top),  $55<\ETfcal<80$ (middle) and $25<\ETfcal<55$ (bottom) for $0.5<\ptab<4$ GeV~\cite{atlaslink}. The recoil contribution is removed via a simple subtraction of the 2D per-trigger yield distribution in a given \ETfcal\ class by that in the peripheral class of $\ETfcal<20$ GeV. The resulting per-trigger yield distribution plus original pedestal in this \ETfcal\ class is then re-normalized over $2<|\Delta\eta|<5$ to obtain the recoil-subtracted 2D correlation function $C_{\Delta}(\Delta\phi,\Delta\eta)$.} 
\end{figure}

The recoil subtraction procedure discussed above is repeated for each $\Deta$ slice, and the resulting 2D or 1D distributions are then converted back into correlation functions, labelled as $C_{\Delta}(\Dphi,\Deta)$ or $C_{\Delta}(\Dphi)$, respectively. The normalization of these distributions are fixed by requiring the average of $C_{\Delta}(\Dphi)$ defined for $2<|\Deta|<5$ to be unity. Figure~\ref{fig:m2a} compares the original 2D correlation functions $C(\Dphi,\Deta)$ (left panels) with the recoil-subtracted correlation functions $C_{\Delta}(\Dphi,\Deta)$ (right panels) in three $\ETfcal$ classes. An away-side ridge is clearly present in all three recoil-subtracted distributions. This away-side ridge is flat to $|\Delta\eta|=5$, with a magnitude similar to that for the near-side ridge. Interestingly, the recoil-subtraction procedure also suppress the near-side jet fragmentation peak around $(\Dphi,\Deta)\sim(0,0)$ to less than 10-15\% of its original height. This implies that the properties of the double-ridge may be extracted for pairs with a modest $\Deta$ gap, as is done by the ALICE collaboration~\cite{ALICE:2012} ($0.8<|\Deta|<1.8$) and PHENIX collaboration~\cite{Adare:2013piz} ($0.5<|\Deta|<0.7$).

\begin{figure}[!h]
\centering
\includegraphics[width=0.7\columnwidth]{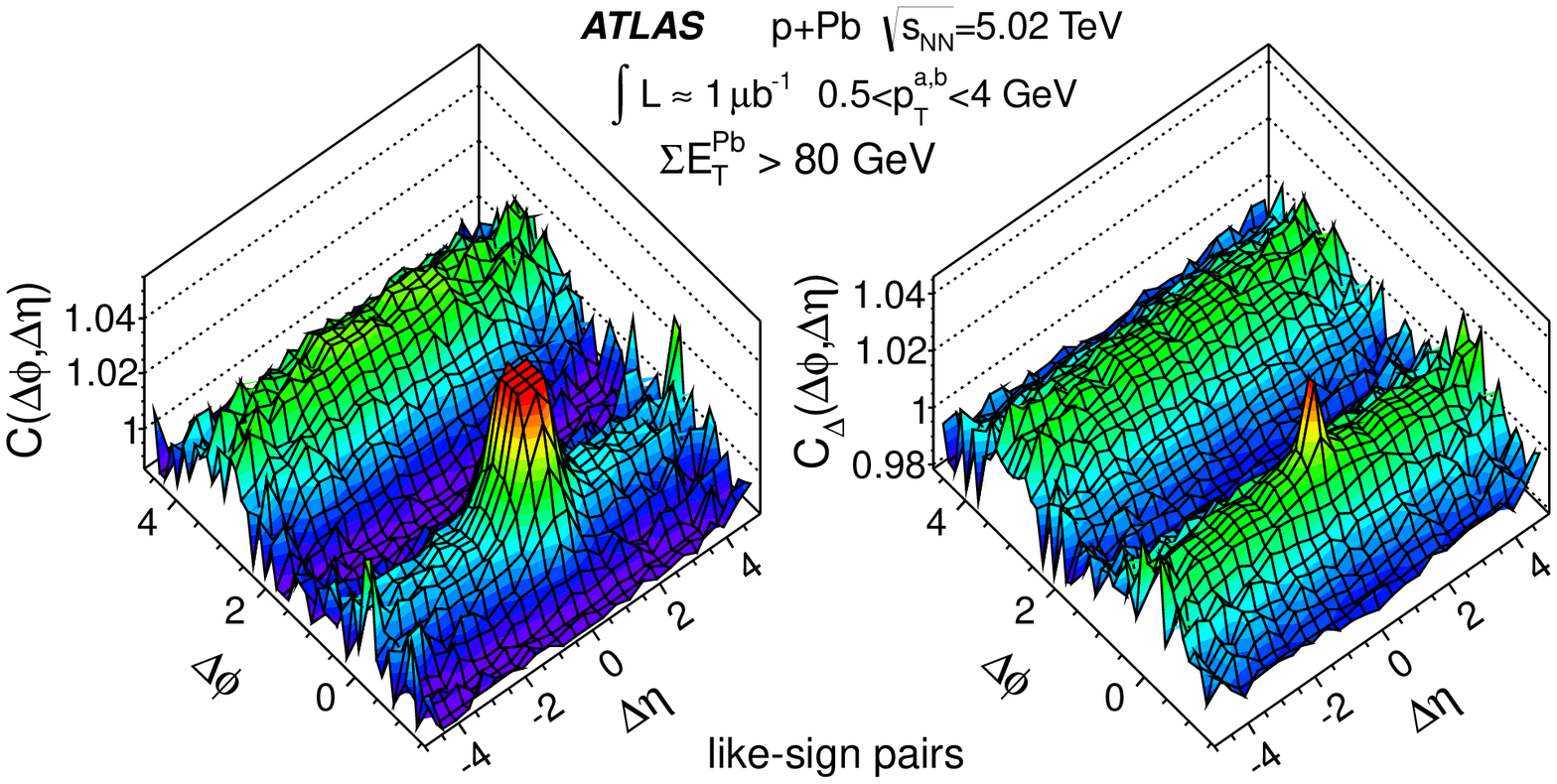}\vspace*{-0.2cm}
\includegraphics[width=0.7\columnwidth]{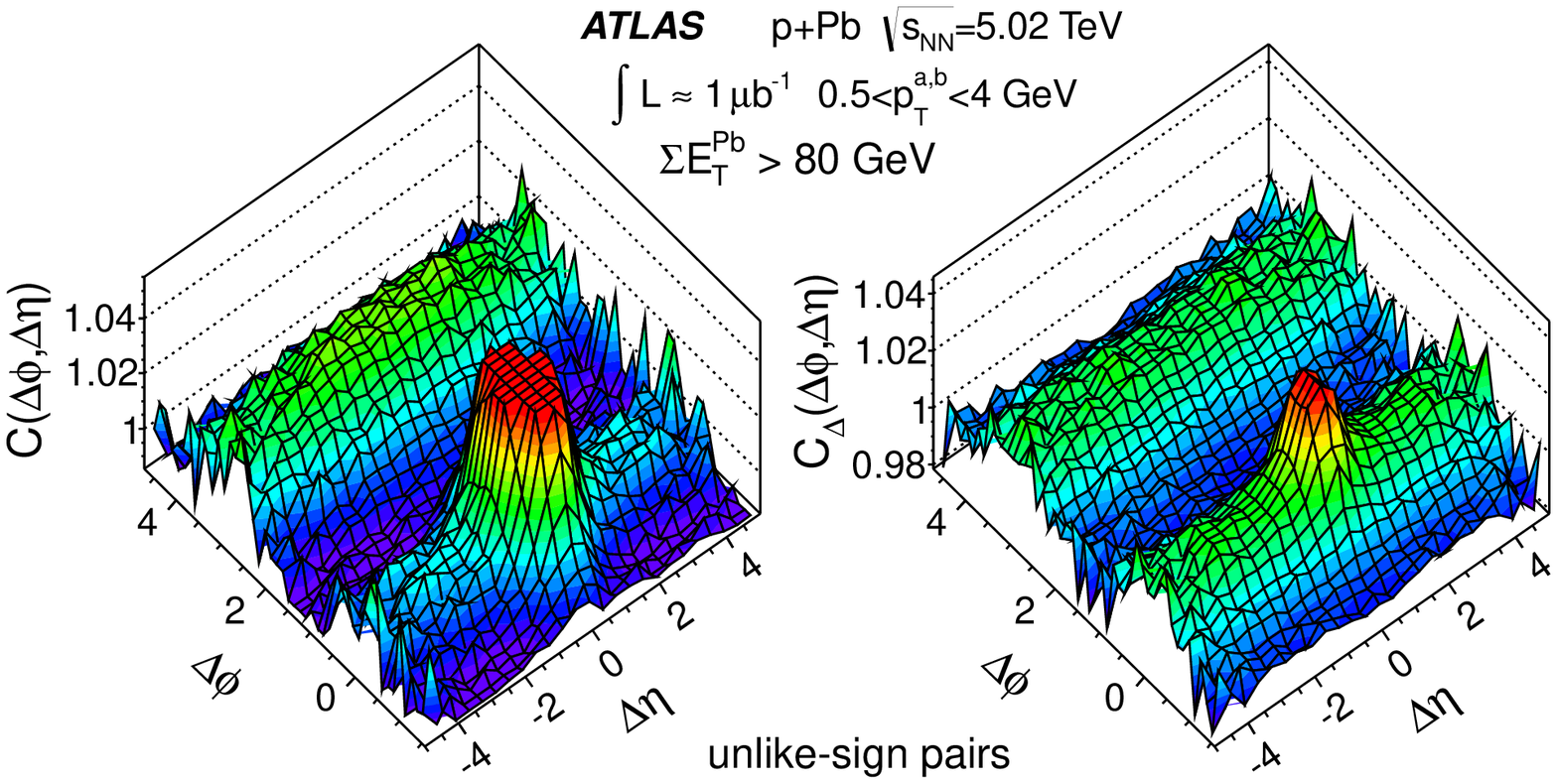}\vspace*{-0.2cm}
\caption{\label{fig:m2b} Two-dimensional correlation function, $C(\Delta\phi,\Delta\eta)$, without (left) and with (right) subtraction of recoil contribution in events with $\ETfcal>80$ GeV and $0.5<\ptab<4$ GeV for like-sign pairs (top) and unlike-sign pairs (bottom)~\cite{atlaslink}.}
\end{figure}

Figure~\ref{fig:m2b} shows the recoil-subtracted 2D correlation functions separately for same-charge pairs and opposite-charge pairs. Despite the large differences in their residual short-range correlations, the extracted double-ridge are nearly identical. 

\begin{figure}[!h]
\centering
\includegraphics[width=0.9\columnwidth]{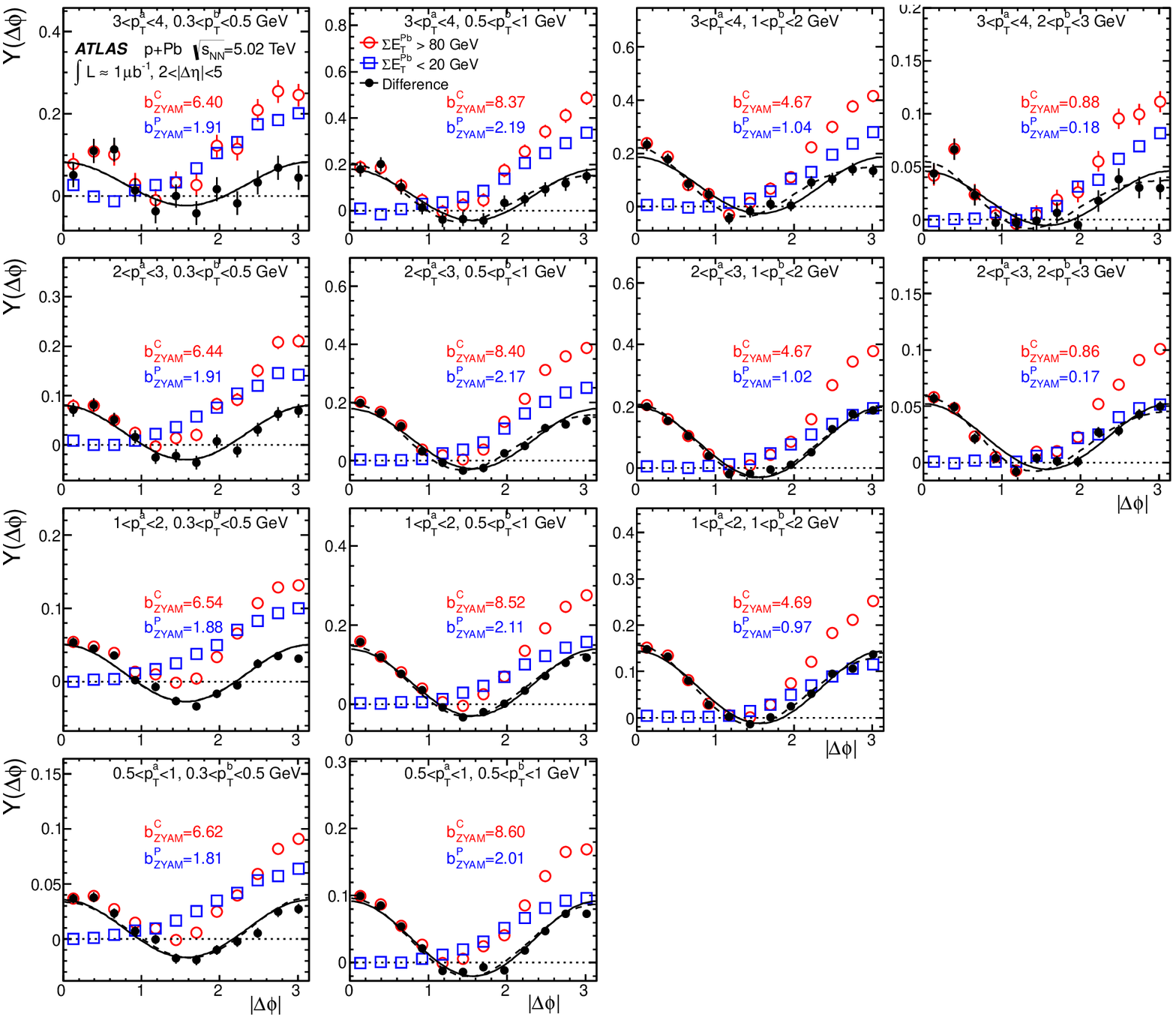}
\caption{\label{fig:m2c}  Distributions of per-trigger yield in the $\ETfcal>80$ GeV event class (open circles), the $\ETfcal<20$ GeV event class (open boxes), and their difference (solid circles) for various ranges of $\ptab$, together with functions, $a_0+2a_2\cos2\Delta\phi$ (solid lines) and $a_0+2a_2\cos2\Delta\phi+2a_3\cos3\Delta\phi$ (dashed lines), obtained via a Fourier decomposition~\cite{atlaslink}.}
\end{figure}

The results discussed so far are obtained for a broad $\pT$ range of 0.5--4 GeV. The double-ridge has also been extracted as a function of $\pT$ as shown in Fig.~\ref{fig:m2c}. The distributions remain largely symmetric around $\Dphi=\pi/2$, however, a significant $\cos 3\Dphi$ component becomes apparent at $\ptab>2-3$ GeV.

\begin{figure}[h]
\centering
\includegraphics[width=0.35\columnwidth]{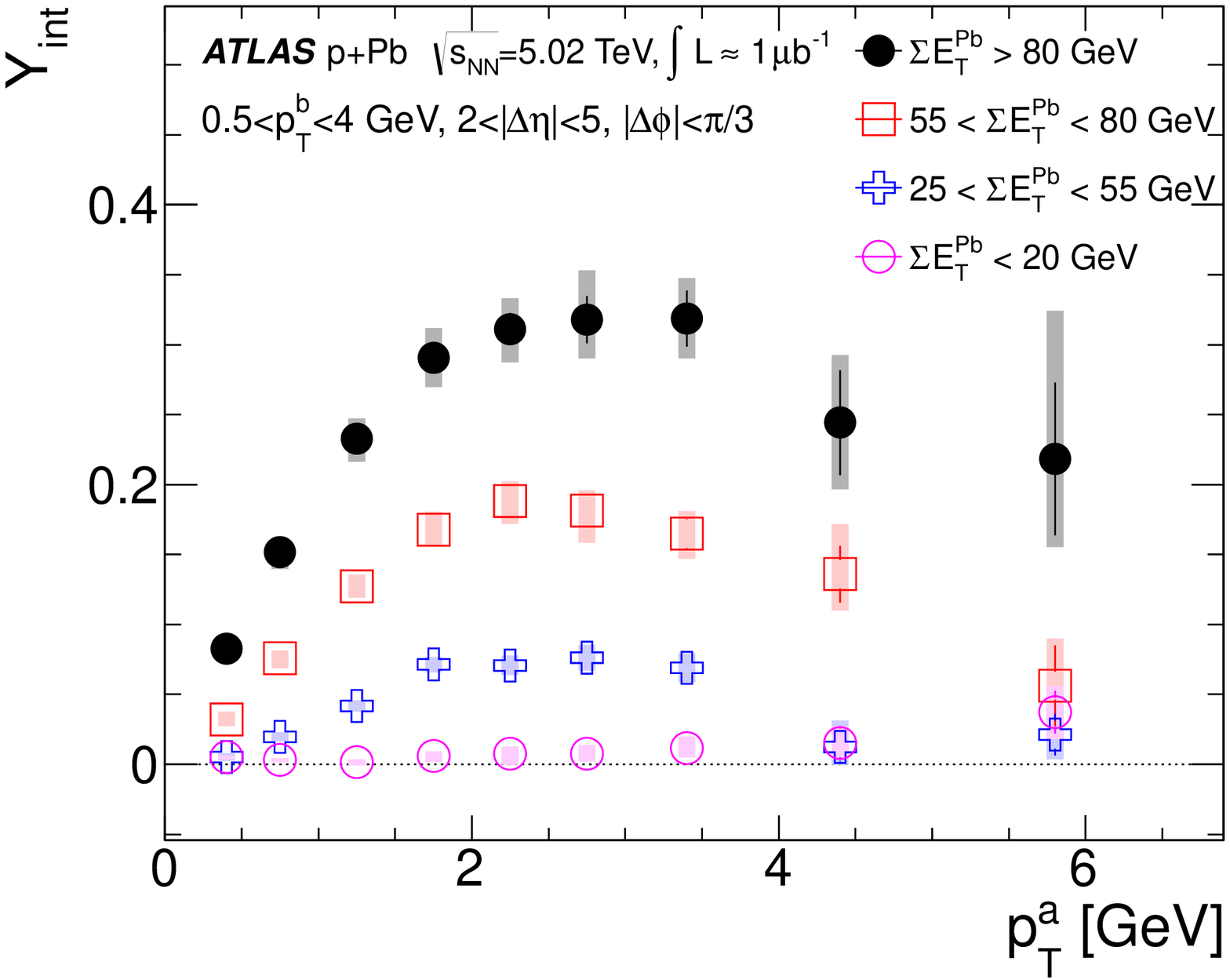}\includegraphics[width=0.35\columnwidth]{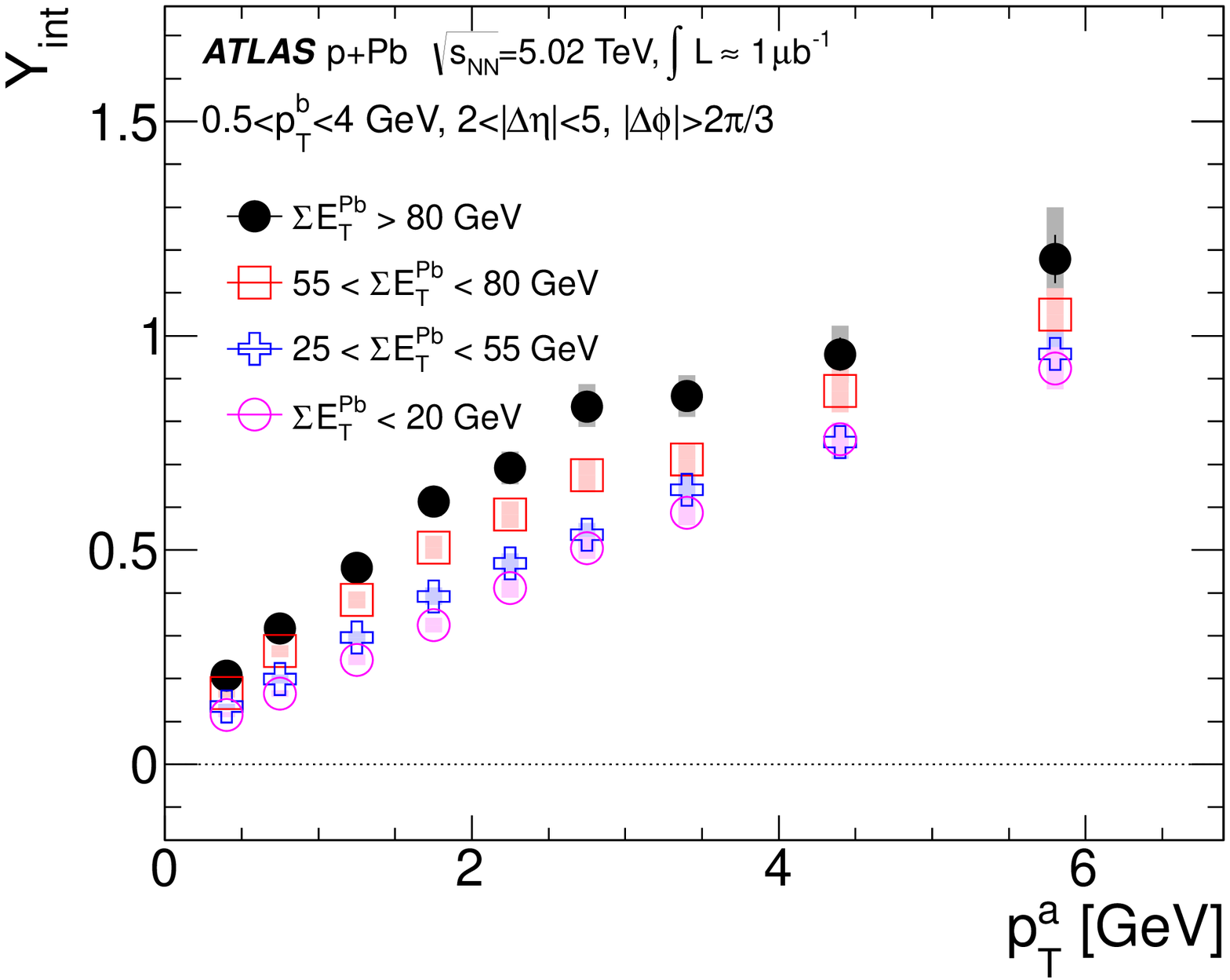}\\
\includegraphics[width=0.35\columnwidth]{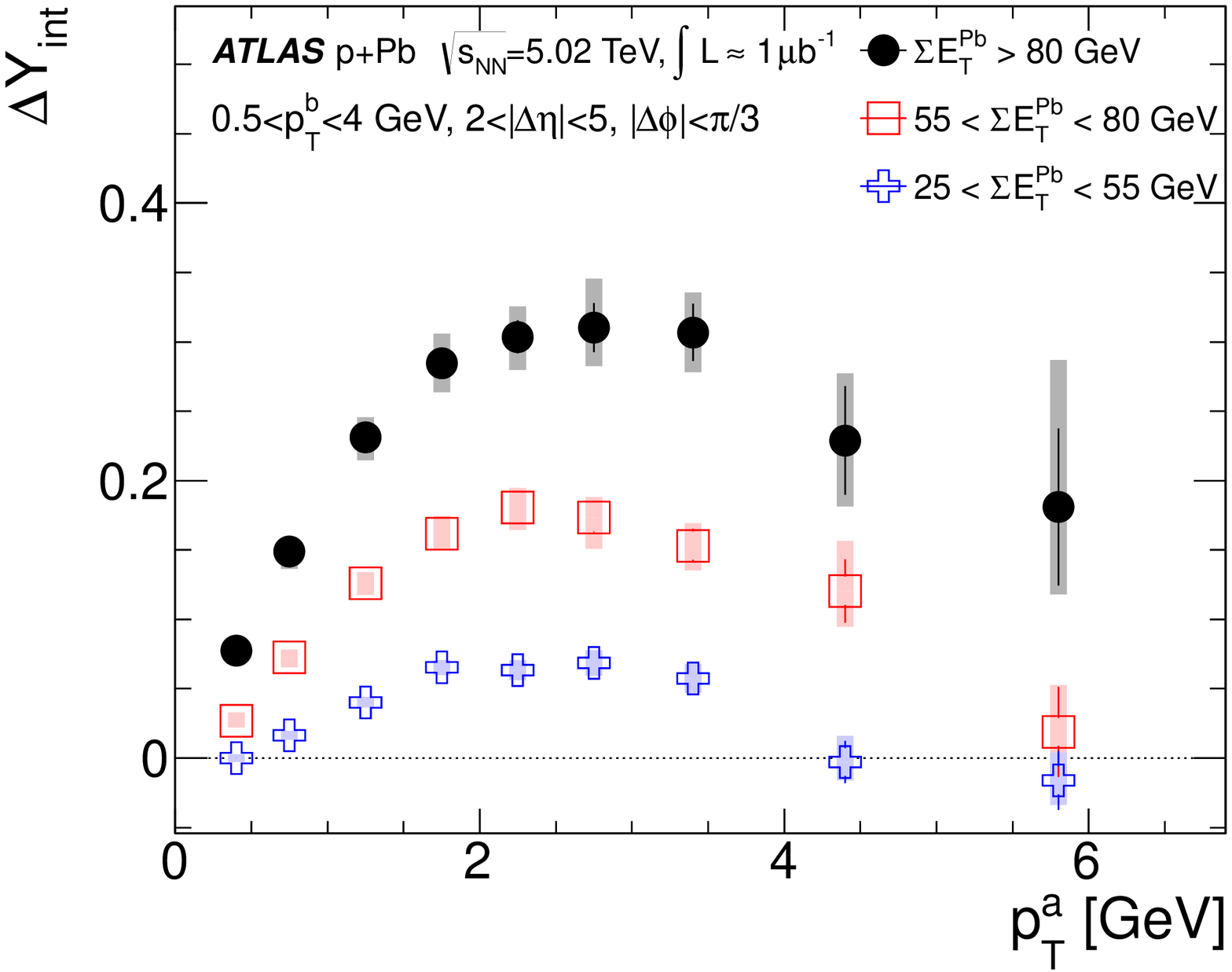}\includegraphics[width=0.35\columnwidth]{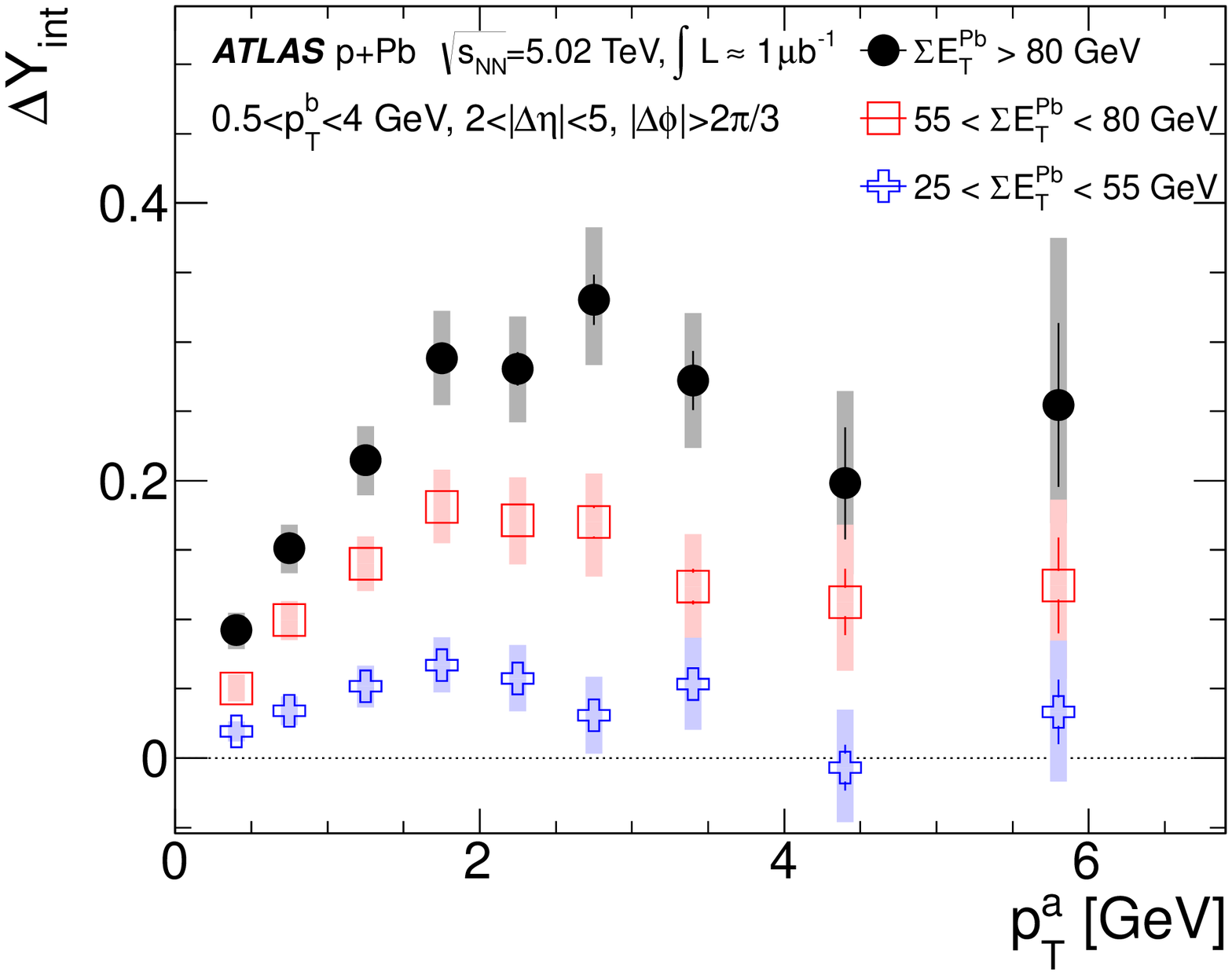}
\caption{\label{fig:m2d} Per-trigger yields vs $\pta$ for $0.5<\ptb<4$~GeV in various $\ETfcal$ event classes on the near-side (top-left panel) and away-side (top-right panel). The bottom panels show the difference of the yield from that in the $\ETfcal<20$ GeV event class~\cite{atlaslink}.}
\end{figure}

To quantify the symmetry between the near- and away-side ridges, the per-trigger yields are integrated over $|\Dphi|<\pi/3$ and $|\Dphi|>2\pi/3$, and plotted in Fig.~\ref{fig:m2d} as a function of $\pta$ with $0.5<\ptb<4$ GeV. According to Eq.~\ref{eq:recoil1}, they directly reflect the $\pT$ dependence of the fractional contribution of the ridge in the correlation function. The differences of the integrated yields between a given $\ETfcal$ class and the $\ETfcal<20$ GeV class (\DelYint) are shown in the bottom panels. The values of $\DelYint$ show a similar magnitude and \pta\ dependence between the near-side and away-side: they rise with \pta\ and reach a maximum around 3--4~GeV. This pattern is visible for the near-side even before subtraction (top-left panel), but is less evident in the away-side before subtraction (top-right panel) due to the dominant contribution of the recoil component. The values of $\DelYint$ increase with $\ETfcal$, but remain symmetric between the near and away-side over the measured $\pta$ range.

\begin{figure}[!h]
\centering
\includegraphics[width=0.35\columnwidth]{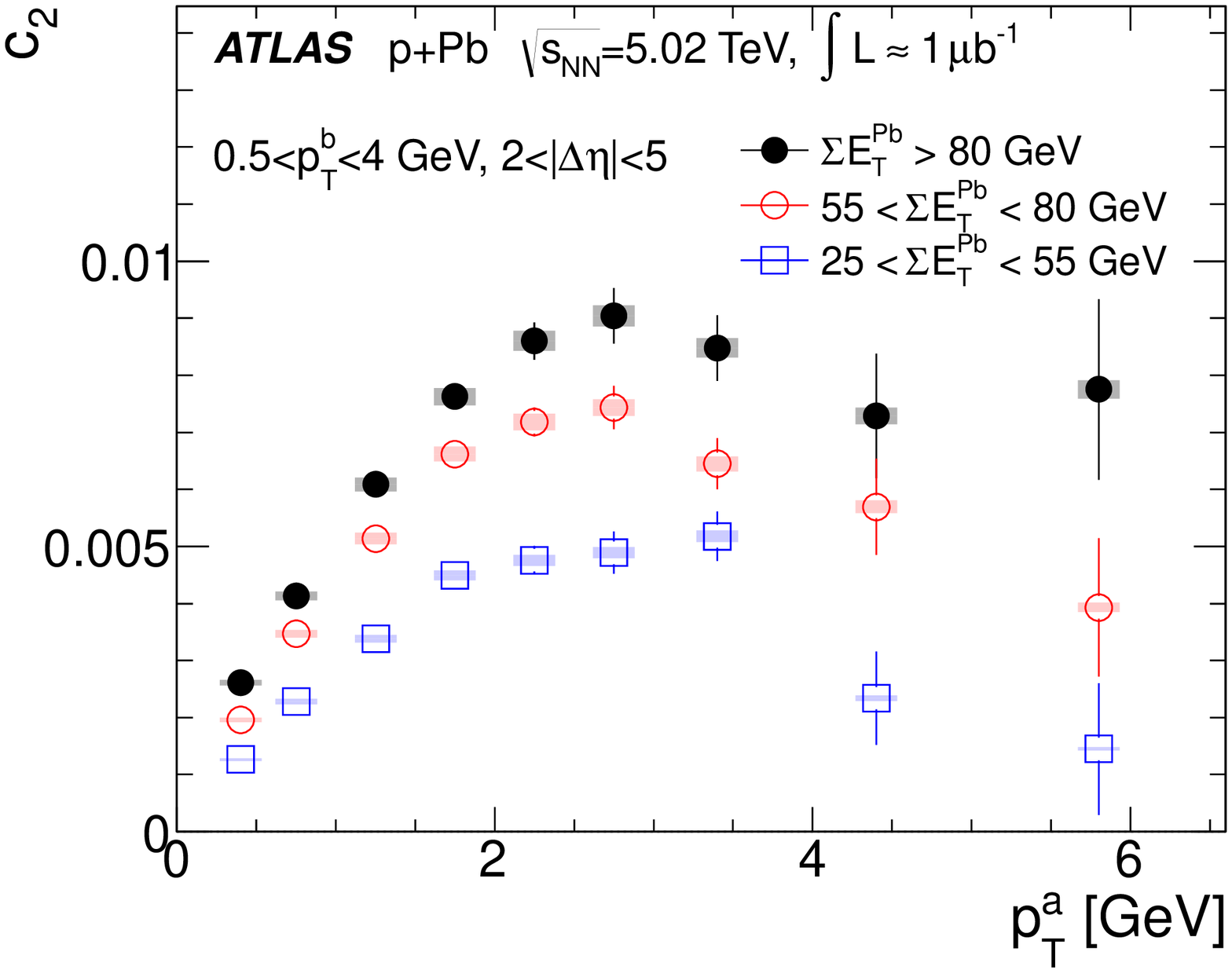}\includegraphics[width=0.35\columnwidth]{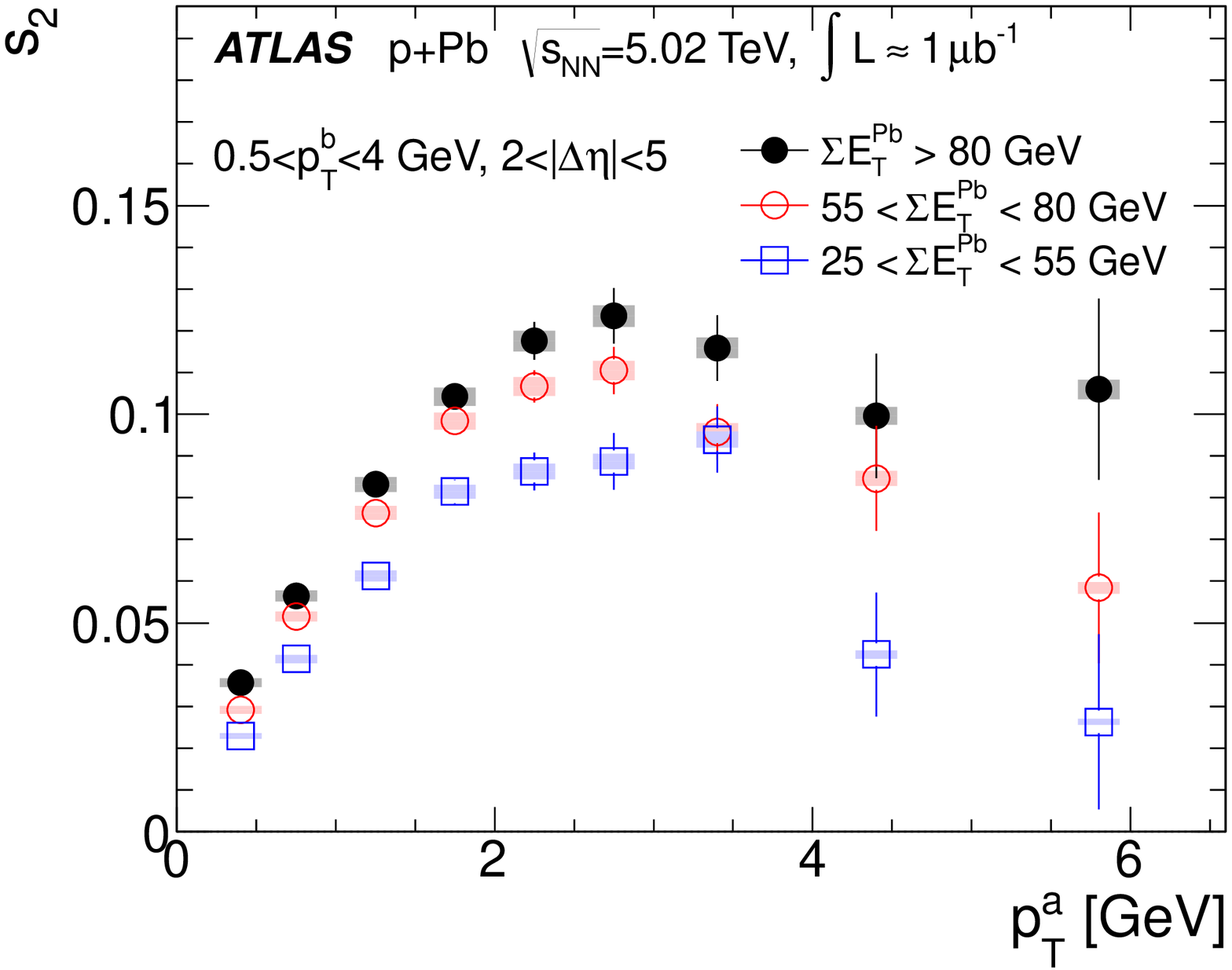}
\includegraphics[width=0.35\columnwidth]{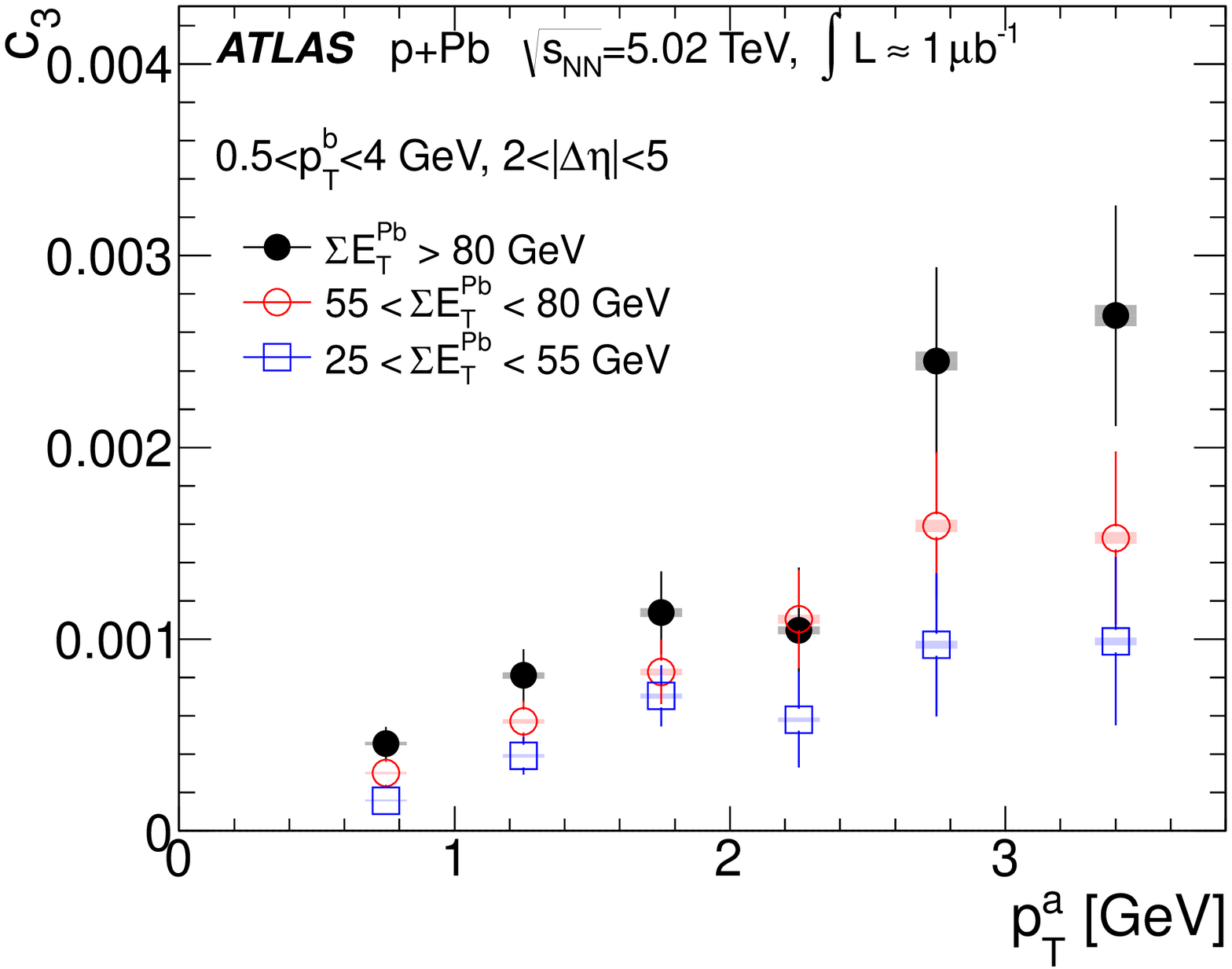}\includegraphics[width=0.35\columnwidth]{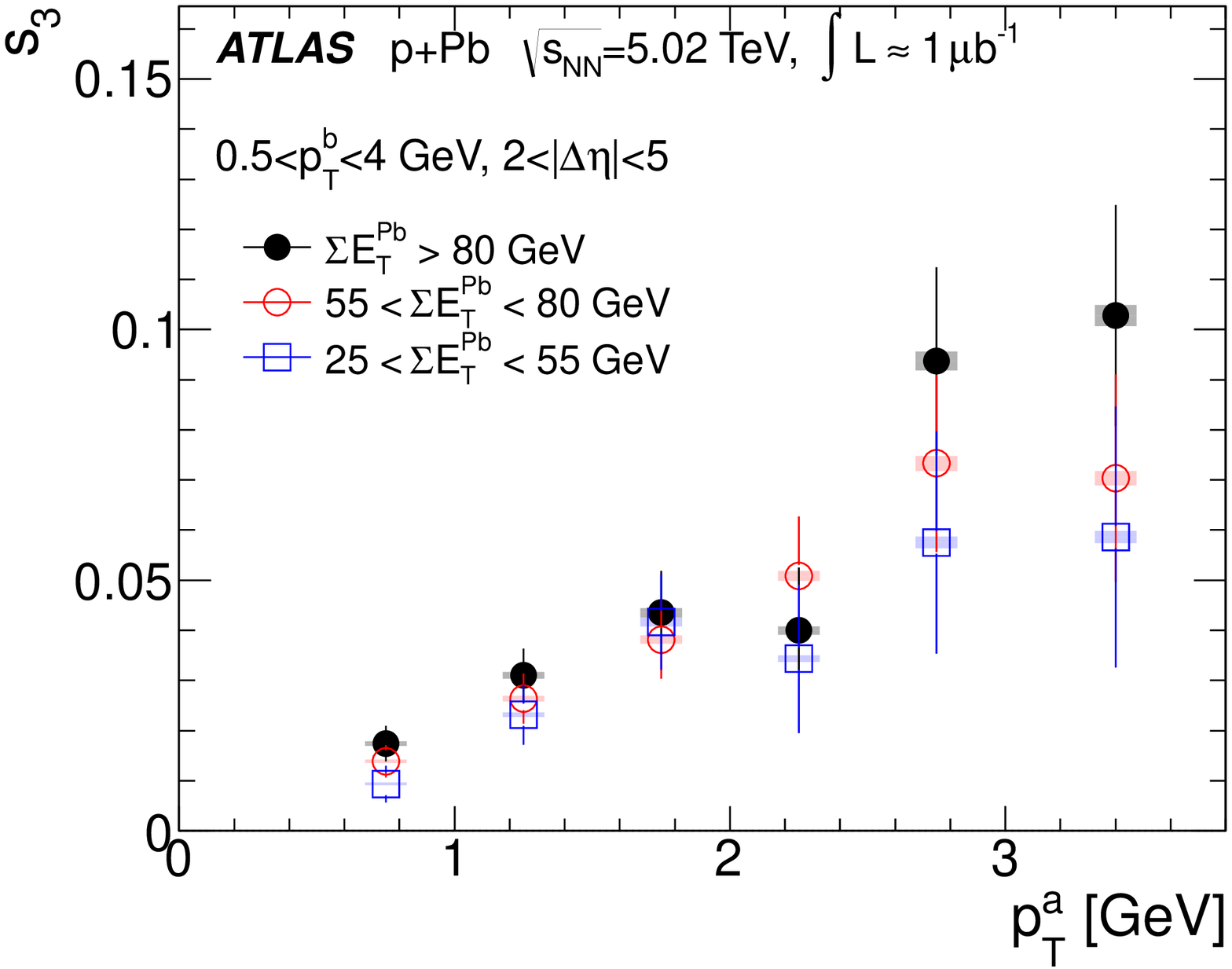}
\caption{\label{fig:m3b} The $c_n$ (left) and $s_n$ (right) vs $\pta$ for $0.5<\ptb<4$~GeV in various \ETfcal\ event classes~\cite{atlaslink}.}
\end{figure}
\section{Harmonic spectra and their factorization behavior}
\label{sec:4}
The amplitude of the $\cos{n\Dphi}$ modulation of $\Delta \Yphi$ relative to the underlying event, $c_n$, for $n=2,3$ is estimated using $a_n$, and the extracted value of $b_{_{\mathrm{ZYAM}}}$ for central events (see Fig.~\ref{fig:m1b}): 
\begin{equation}
c_n = a_n/(b^{{\mathrm{C}}}_{_{\mathrm{ZYAM}}} + a_0).
\label{eq:ctwo}
\end{equation}
Equivalantly, it can also be calculated as the Fourier coefficients of the recoil-subtracted correlation function $C_\Delta(\Dphi)$: $c_n = \langle C_\Delta(\Dphi)\cos n\Dphi\rangle$.

Figure~\ref{fig:m3b} shows $c_2$ (top-left panel) and $c_3$ (bottom-left panel) as a function of \pta\ for $0.5<\ptb<4$~GeV in three $\ETfcal$ classes. The values of $c_2$ are much larger than those for $c_3$. The $\pta$ dependence of $c_2$ is similar to that for the $\Delta\Yphi$ in Fig.~\ref{fig:m2d}. The magnitude of the $c_n$ is observed to increase for larger $\ETfcal$, opposite to the predictions from hydrodynamic calculations~\cite{Bozek:2011if,Bzdak:2013zma}. Using the techniques discussed in Ref.~\cite{ATLAS:2012at}, $c_n$ is converted into an estimate of $s_n$, the average $n^{\mathrm {th}}$-order Fourier coefficient of the event-by-event single-particle $\phi$ distribution, by assuming the factorization relation:
\begin{equation}
\label{eq:fac}
c_n(\pta,\ptb)=s_n(\pta)s_n(\ptb).
\end{equation}
From this, $s_n(\pta)$ is calculated as $s_n(\pta)=c_n(\pta,\ptb)/\sqrt{c_n(\ptb,\ptb)}$, where $c_n(\ptb,\ptb)$ is obtained from $a_n$ calculated from Fig.~\ref{fig:m1b}. The $s_2(\pta)$ values obtained this way exceed 0.1 at $\pT \sim$ 2--4~GeV, and the $s_3(\pta)$ values remain less than 60\% of the $s_2(\pta)$ values over the measured $\pT$ range.

\begin{figure}[!h]
\centering
\includegraphics[width=0.5\columnwidth]{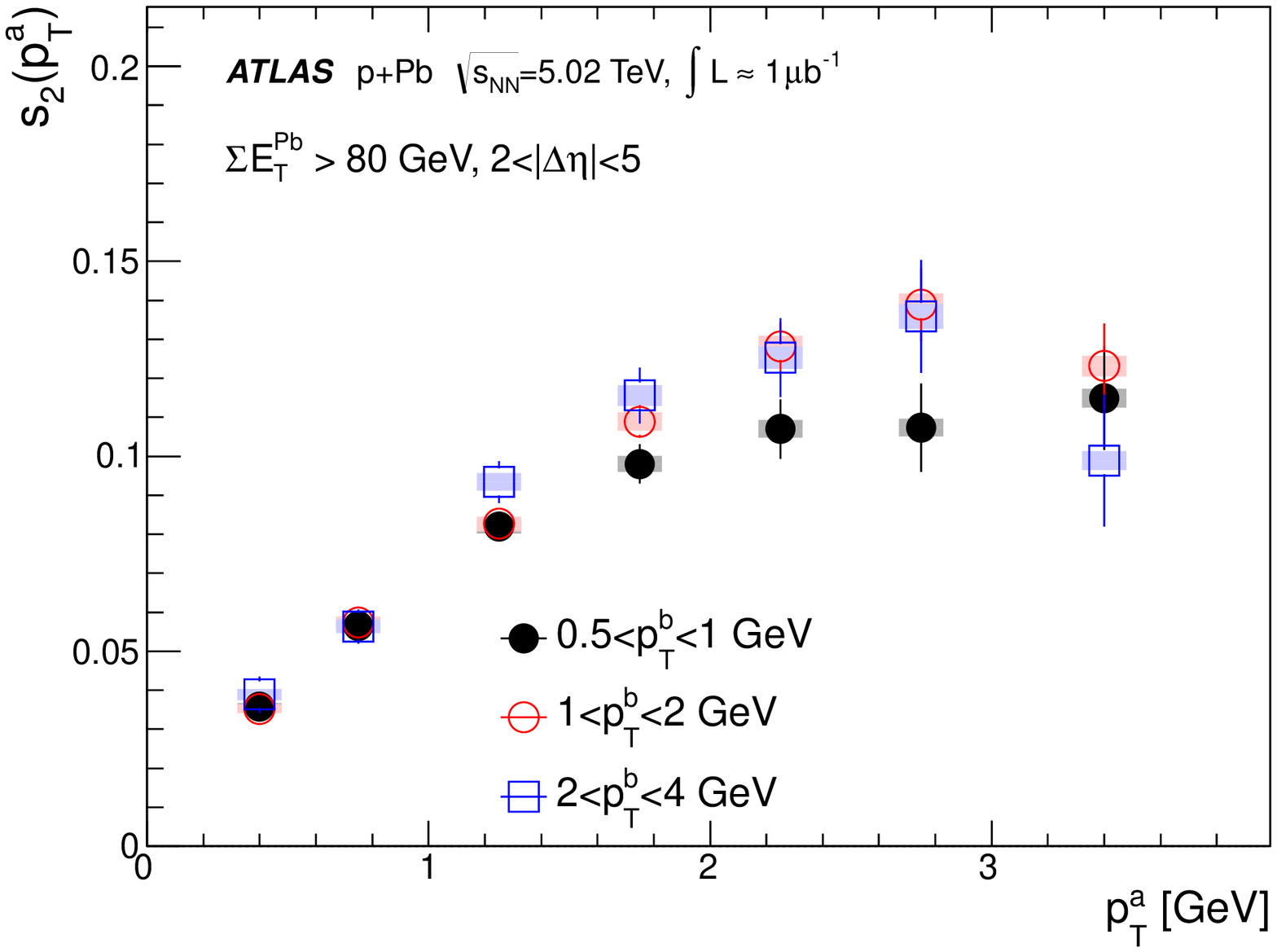}\includegraphics[width=0.5\columnwidth]{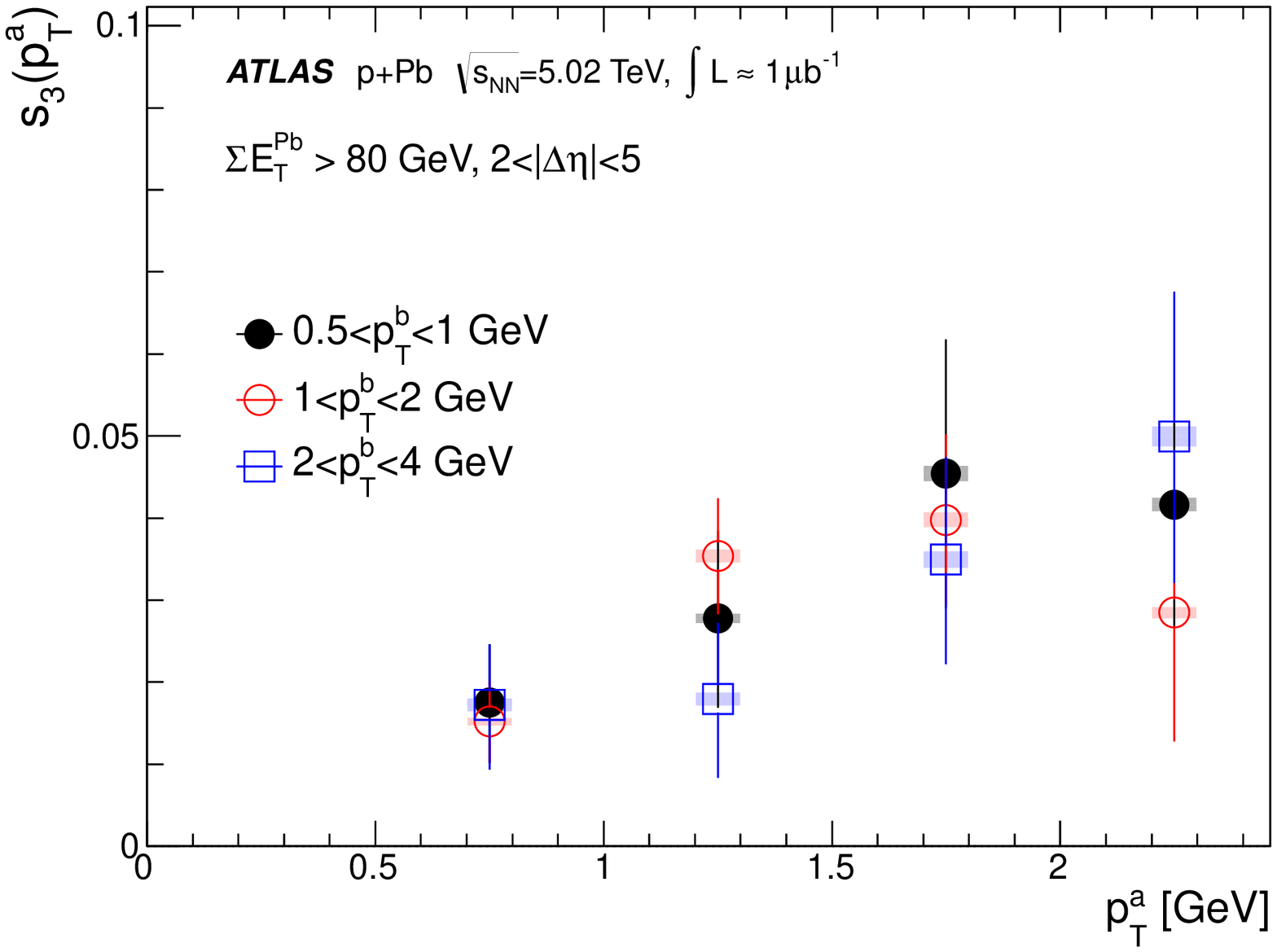}
\caption{\label{fig:m3c} The $s_n(\pta)$ in several $\ptb$ ranges. It quantifies the factorization relation $c_n(\pta,\ptb) = s_n(\pta)s_n(\ptb)$, i.e. checking how well $s_n(\pta)$ extracted for different $\ptb$ ranges agree with each other~\cite{atlaslink}.}
\end{figure}

The factorization relation used to compute $s_n(\pta)$ is checked directly in Fig.~\ref{fig:m3c} using three different sub-ranges of $\ptb$ within 0.5--4~GeV. The factorization is found to be valid within 10\%--20\% for $s_2(\pta)$ , while the precision of $s_3(\pta)$ data does not allow a quantitative test of the factorization. The analysis is also repeated for correlation functions separately constructed from like-sign pairs and unlike-sign pairs, and the resulting $c_n$ and $s_n$ coefficients are found to be consistent within their statistical and systematic uncertainties (see Fig. 6 and Fig. 9 at \cite{atlaslink}). 

\section{Comment on event selection bias}
\label{sec:5}
The event class in the ATLAS analysis is defined by the $\ETfcal$ in $3.1<\eta<4.9$, well separated from the range $|\eta|<2.5$ used to construct the correlation functions. The analysis is also repeated for event classes defined explicitly using the number of reconstructed charged particles $\nchrec$ in $\pT>0.4$ GeV and $|\eta|<2.5$. As shown in Fig.~\ref{fig:cent}, the correlation between $\ETfcal$ and $\nchrec$ is quite broad. The events selected in a $\ETfcal$ interval cover a broad range in $\nchrec$, and vice versa.

\begin{figure}[!h]
\centering
\includegraphics[width=0.33\columnwidth]{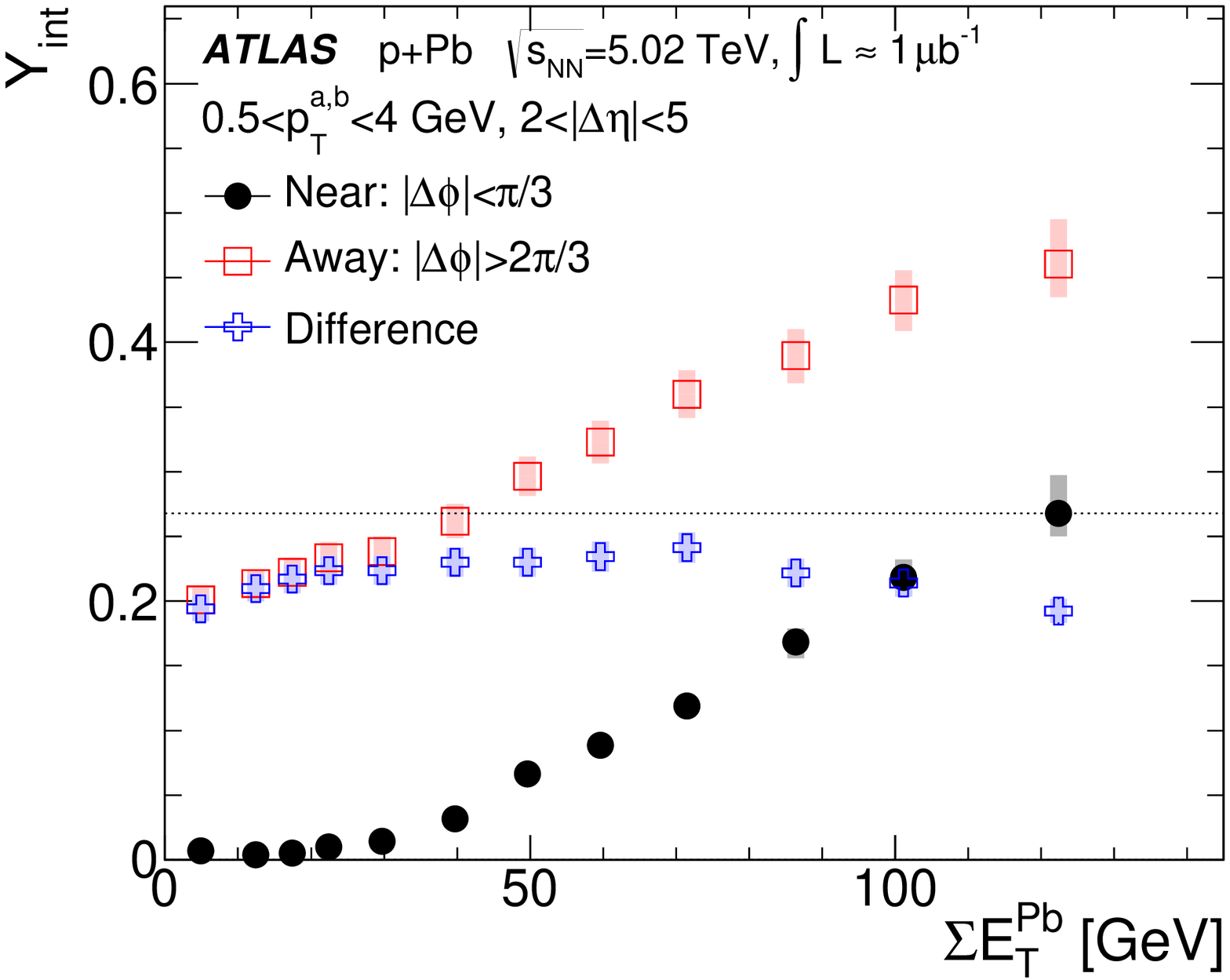}\includegraphics[width=0.33\columnwidth]{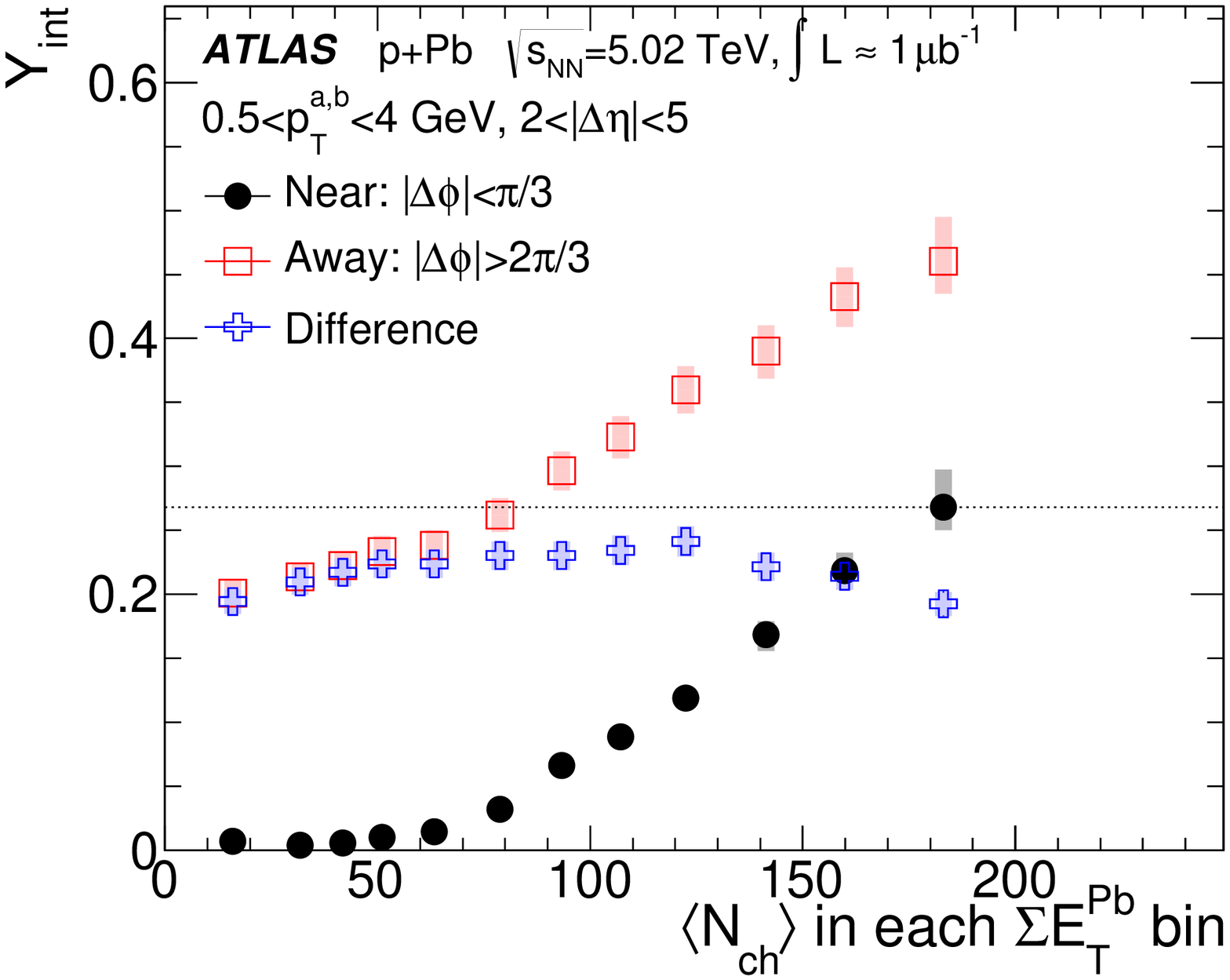}\includegraphics[width=0.33\columnwidth]{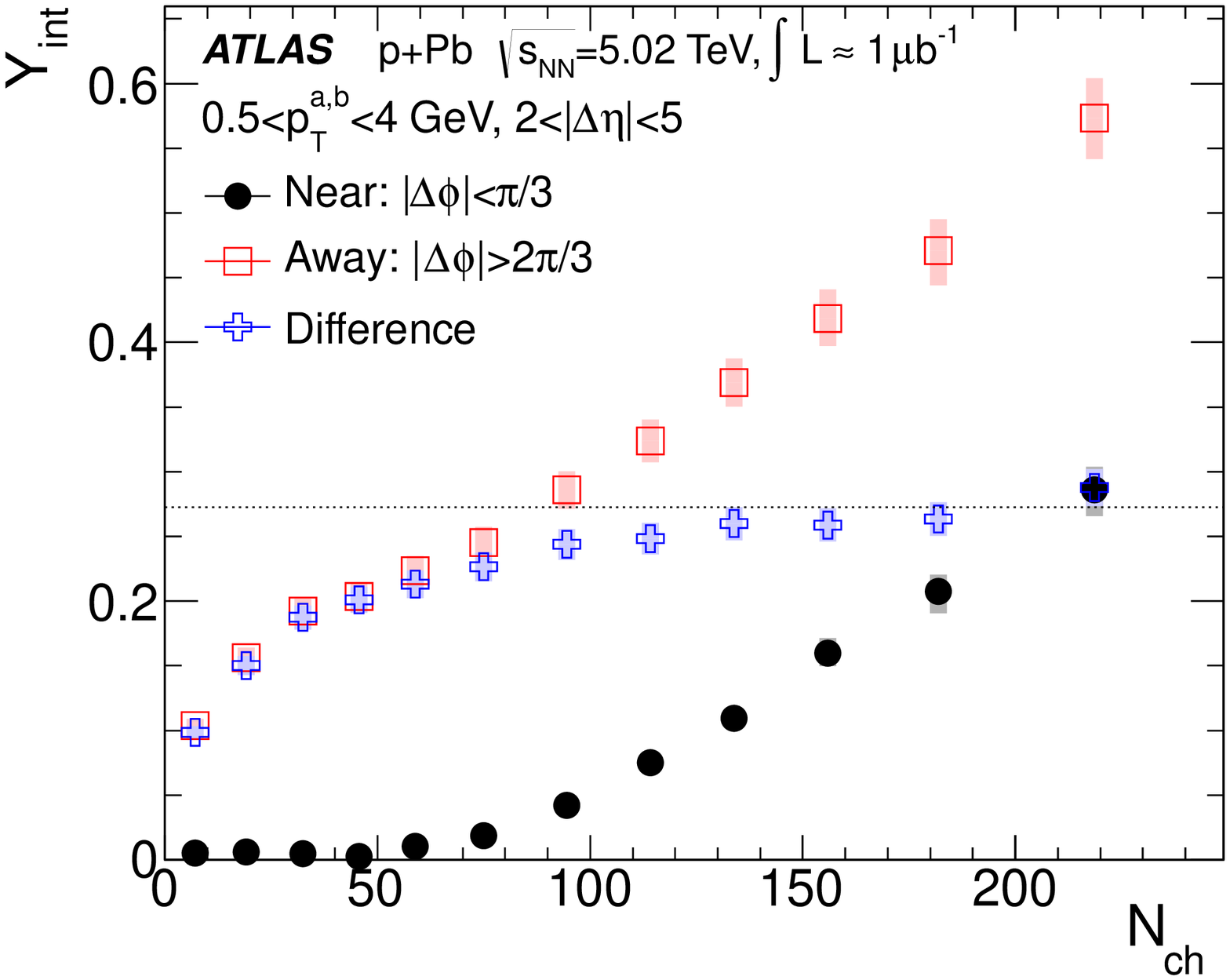}\vspace*{-0.2cm}
\caption{\label{fig:m4b} Integrated per-trigger yields for $0.5<\ptab<4$~GeV as a function of \ETfcal\ for events selected on \ETfcal\ (left), as a function of \nchave\ for events selected on \ETfcal\ (middle) and as a function of \nch\ for events selected on \nch\ (right). The middle panel is same as left panel except for a change of $x$-axis using the \nchave\ given in Table~\ref{tab:1}. The away-side yield in the right panel shows a clear drop for events explicitly required to have small $\nch$, and a stronger increase for events explicitly required to have large $\nch$, indicating the presence of auto-correlation bias~\cite{atlaslink}.}\vspace*{-0.2cm}
\end{figure}

Figure~\ref{fig:m4b} compares the integrated per-trigger yield as a function of $\ETfcal$ for events selected via $\ETfcal$ (left panel), as a function of $\nchave$ for events selected via $\ETfcal$ (middle panel) and as a function of $\nch$ for events selected via $\nchrec$ (right panel). The near-side ridge yields in the three cases show very similar trends and reach nearly the same value ($\sim$0.26) in the most central bin, suggesting that the three results are related to each other by an overall scale factor in the $x$-axis. The away-side yield, however, show quite different dependence between the right panel and the left two panels. In particular, the bend-over behavior at small $\nch$ and the increase at large $\nch$ can be attributed to auto-correlation bias associated with explicit selection of events based on $\nch$: events required to have small number of $\nch$ preferably select events containing jets with smaller number of fragments, while events with large $\nch$ preferably select events containing jets with larger number of fragments. This bias can be significantly reduced if the yields are presented as a function of $\ETfcal$ or $\nchave$ for events classes based on the $\ETfcal$ (left and middle panels).

\begin{figure}[!h]
 \centering
 \includegraphics[width=1\columnwidth]{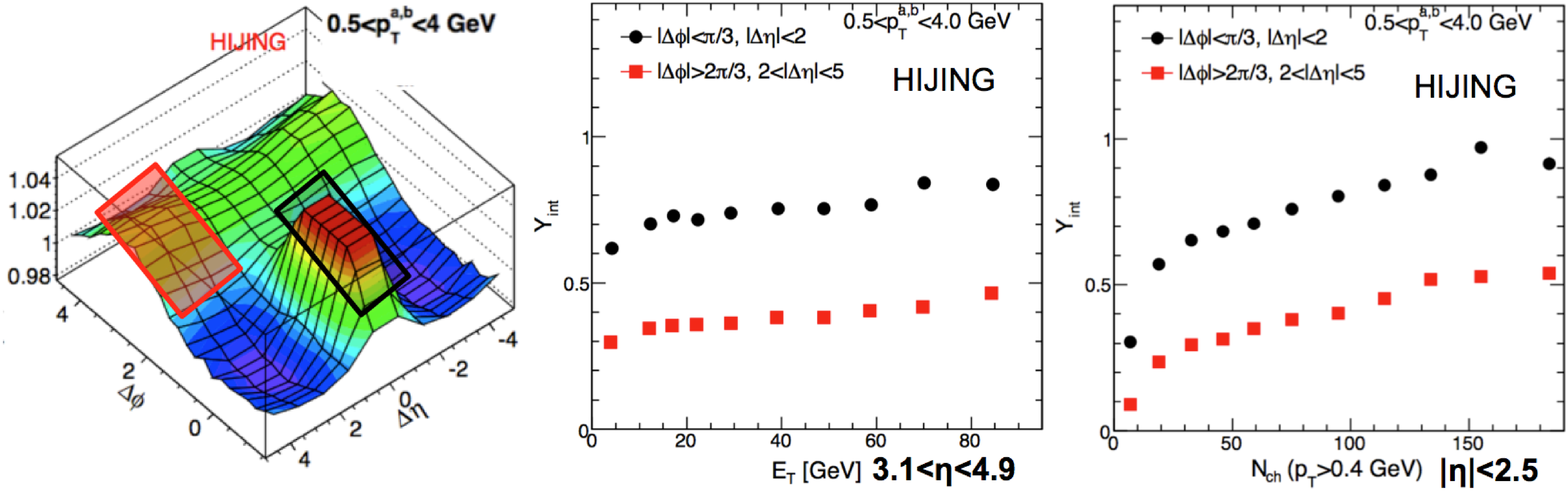}
 \caption{\label{fig:m4c} Illustration of event selection bias via HIJING simulation: (left) 2D correlation function and the two integration regions for per-trigger yield marked by the boxes, (middle) the yield vs forward $E_T$ in $3.1<\eta<4.9$ for events selected with  $E_T$, (right) the yield vs $\nch$ in $|\eta|<2.5$ for events selected on $\nch$.}\vspace*{-0.2cm}
 \end{figure}

This bias is checked explicitly using HIJING simulation as shown in Fig.~\ref{fig:m4c}. HIJING has no flow physics, so all the structures in the 2D correlation function can be attributed to correlations involving a subset of the particles in the events. The per-trigger yields as a function of $E_{\mathrm T}$ over $3.1<\eta<4.9$ range show very weak variation, while the yields as a function of $\nch$ for events selected on $\nch$ show much stronger variation, very similar to the observation in the data.

\section{Summary}
\label{sec:6}
ATLAS has measured two-particle correlations in $\sqrtsnn =5.02$~TeV $\pPb$ collisions in broad ranges of event activity (quantified by $\ETfcal$, the $E_{\mathrm T}$ in $3.2<\eta<4.9$ in Pb-going side) and $\pT$ (0.5-7 GeV). After subtracting the contributions from jet fragmentation around $(\Dphi,\Deta)\sim(0,0)$ and $\Dphi \sim\pi$, the resulting correlations show a significant long-range correlation that extends to $|\Deta|=5$ and are approximately symmetric around $\Dphi=\pi/2$. The $\Dphi$-shape of these correlations is well described by a $1+2c_2\cos2\Dphi +2c_3\cos3\Dphi$ function over the measured ranges in $\pT$, $\Deta$ and $\ETfcal$. The values of $c_2$ are found to increase with $p_{\mathrm T}$ to 3-4 GeV and fall at higher $p_{\mathrm T}$, the values of the $c_3$ are found to be negligible at low $p_{\mathrm T}$ but increase to about 30\% of the $c_2$ at high $p_{\mathrm T}$. The overall magnitude of the $c_n$ decreases by about 50\% between events with $\ETfcal>80$ and $25<\ETfcal<55$ GeV. The $c_n$ parameters are converted to the Fourier coefficients $s_n$ for single-particle azimuthal distribution using a factorization ansatz Eq.~\ref{eq:fac}. The magnitude of the extracted $s_2$ ($s_3$) reaches more than 0.1 (0.05) in 3-4 GeV for $\ETfcal>80$ GeV, but is about 20\% lower for events with $25<\ETfcal<55$ GeV. These findings are consistent with final-state collective effects in high-multiplicity events~\cite{Bozek:2011if,Shuryak:2013ke,Bozek:2013uha}, but they are also compatible with predictions in the Color Glass Condensate approach~\cite{Dusling:2012wy,Dusling:2013oia}.

This proceeding also addresses several technical aspects of the 2PC analysis, including the validity of the recoil-subtraction procedure and auto-correlation bias associated with the definition of event activity class. The recoil-subtraction procedure is demonstrated to work well (10-15\%) in the relatively low $\pT$ range of this analysis ($\pT<7$ GeV) where ridge structure is dominant. The definition of event activity class using $\ETfcal$ in rapidity range ($3.1<\eta<4.9$) not overlapping with particles used in the 2PC analysis ($|\eta|<2.5$), is shown to reduce the auto-correlation bias. In contrast, this bias is clearly present when the particles involved in the correlation analysis are also used to define the event classes.

This work is in part supported by NSF under grant number PHY-1019387.
\section*{References}
\bibliography{JiangyongJia}{}

\providecommand{\newblock}{}
\begin{thebibliography}{10}
\expandafter\ifx\csname url\endcsname\relax
  \def\url#1{{\tt #1}}\fi
\expandafter\ifx\csname urlprefix\endcsname\relax\def\urlprefix{URL }\fi
\providecommand{\eprint}[2][]{\url{#2}}

\bibitem{Abelev:2009af}
{STAR Collaboration} 2009 {\em Phys.~Rev.~C\/} {\bf 80} 064912

\bibitem{Alver:2009id}
{PHOBOS Collaboration} 2010 {\em Phys.Rev.Lett.\/} {\bf 104} 062301

\bibitem{Adare:2008cqb}
{PHENIX Collaboration} 2008 {\em Phys.~Rev.~C\/} {\bf 78} 014901

\bibitem{ATLAS:2012at}
{ATLAS Collaboration} 2012 {\em Phys.~Rev.~C\/} {\bf 86} 014907

\bibitem{Aamodt:2011by}
{ALICE Collaboration} 2012 {\em Phys.~Lett.\/} {\bf B708} 249

\bibitem{Chatrchyan:2012wg}
{CMS Collaboration} 2012 {\em Eur.~Phys.~J.\/} {\bf C72} 2012

\bibitem{Khachatryan:2010gv}
{CMS Collaboration} 2010 {\em JHEP\/} {\bf 09} 091

\bibitem{CMS:2012qk}
{CMS Collaboration} 2013 {\em Phys.Lett.\/} {\bf B718} 795

\bibitem{ALICE:2012}
{ALICE Collaboration} 2013 {\em Phys.Lett.\/} {\bf B719} 29

\bibitem{Aad:2012gla}
{ATLAS Collaboration} 2013 {\em Phys.Rev.Lett.\/} {\bf 110} 182302

\bibitem{Voloshin:2008dg}
Voloshin S~A, Poskanzer A~M and Snellings R  \textit{Preprint}
  \eprint{0809.2949}

\bibitem{Bozek:2011if}
Bozek P 2012 {\em Phys.Rev.\/} {\bf C85} 014911

\bibitem{Bozek:2012gr}
Bozek P and Broniowski W 2013 {\em Phys.Lett.\/} {\bf B718} 1557

\bibitem{Shuryak:2013ke}
Shuryak E and Zahed I  \textit{Preprint} \eprint{1301.4470}

\bibitem{Bozek:2013uha}
Bozek P and Broniowski W  \textit{Preprint} \eprint{1304.3044}

\bibitem{Dusling:2012wy}
Dusling K and Venugopalan R 2012 {\em Phys.~Rev.~D\/} {\bf 87} 054014

\bibitem{Dusling:2013oia}
Dusling K and Venugopalan R  \textit{Preprint} \eprint{1302.7018}

\bibitem{atlaslink}
ATLAS Collaboration,
  https://atlas.web.cern.ch/Atlas/GROUPS/PHYSICS/PAPERS/HION-2012-13/

\bibitem{Aad:2008zzm}
{ATLAS Collaboration} 2008 {\em JINST\/} {\bf 3} S08003

\bibitem{Ajitanand:2005jj}
Ajitanand N {\em et~al.\/} 2005 {\em Phys.~Rev.~C\/} {\bf 72} 011902

\bibitem{Borghini:2000cm}
Borghini N, Dinh P~M and Ollitrault J~Y 2000 {\em Phys.Rev.\/} {\bf C62} 034902

\bibitem{Borghini:2002mv}
Borghini N, Dinh P, Ollitrault J~Y, Poskanzer A~M and Voloshin S 2002 {\em
  Phys.Rev.\/} {\bf C66} 014901

\bibitem{Adare:2013piz}
{PHENIX Collaboration} 2013  \textit{Preprint} \eprint{1303.1794}

\bibitem{Bzdak:2013zma}
Bzdak A, Schenke B, Tribedy P and Venugopalan R  \textit{Preprint}
  \eprint{1304.3403}

\end{thebibliography}
\bibliographystyle{iopart-num}

\end{document}